\documentclass[acmlarge]{acmart}
\usepackage{balance}       
\usepackage{graphics}      
\usepackage{hyperref}
\usepackage{color}
\usepackage{booktabs}
\usepackage{textcomp}
\usepackage{subcaption}
\usepackage{enumerate}
\usepackage{xcolor}
\usepackage{lipsum}
\usepackage{makecell}
\usepackage{multicol}
\usepackage{multirow}
\usepackage{array}
\usepackage{verbatimbox}
\usepackage{enumitem}
\usepackage{amsmath}
\usepackage{stfloats}
\usepackage{graphicx}
\usepackage{amsthm}
\usepackage{listings}
\usepackage{caption} 
\usepackage[export]{adjustbox}
\usepackage{xspace}
\usepackage{epsfig}
\usepackage[linesnumbered]{algorithm2e}
\usepackage{algpseudocode}
\usepackage{tabularx}
\usepackage{arydshln}
\usepackage[bottom]{footmisc}
\usepackage{tcolorbox}
\usepackage{stackengine}
\usepackage{placeins}
\usepackage{graphicx}

\newcommand*{\eg}{\textit{e.g.},\xspace}
\newcommand*{\ie}{\textit{i.e.},\xspace}
\newcommand*{\vs}{\textit{vs.}\xspace}

\newcolumntype{L}[1]{>{\raggedright\let\newline\\\arraybackslash\hspace{0pt}}m{#1}}
\newcolumntype{C}[1]{>{\centering\let\newline\\\arraybackslash\hspace{0pt}}m{#1}}
\newcolumntype{R}[1]{>{\raggedleft\let\newline\\\arraybackslash\hspace{0pt}}m{#1}}

\makeatletter
\def\thickhline{%
  \noalign{\ifnum0=`}\fi\hrule \@height \thickarrayrulewidth \futurelet
   \reserved@a\@xthickhline}
\def\@xthickhline{\ifx\reserved@a\thickhline
               \vskip\doublerulesep
               \vskip-\thickarrayrulewidth
             \fi
      \ifnum0=`{\fi}}
\makeatother

\makeatletter
\def\thickhlinespace{%
  \addlinespace[1ex]
  \noalign{\ifnum0=`}\fi\hrule \@height \thickarrayrulewidth \futurelet
   \reserved@a\@xthickhline
   \addlinespace[1ex]
   }
\def\@xthickhlinespace{\ifx\reserved@a\thickhline
               \vskip\doublerulesep
               \vskip-\thickarrayrulewidth
             \fi
      \ifnum0=`{\fi}}
\makeatother

\newlength{\thickarrayrulewidth}
\newlength\Origarrayrulewidth

\algnewcommand{\IfThenElse}[3]{
  \State \algorithmicif\ #1\ \algorithmicthen\ #2\ \algorithmicelse\ #3}
  
\newenvironment{s_itemize}{
\begin{itemize}[leftmargin=*]
  \setlength{\itemsep}{3pt}
  \setlength{\parskip}{0pt}
  \setlength{\parsep}{0pt}
}{\end{itemize}}

\definecolor{downredcolor}{HTML}{e31a1c}
\definecolor{upgreencolor}{HTML}{33a02c}

\definecolor{DarkGreen}{HTML}{5DAC81}

\newcommand\projectname{WatchGuardian\xspace}

\begin{document}

%

\title{\projectname: Enabling User-Defined Personalized Just-in-Time Intervention on Smartwatch}

%

\author{Ying Lei}
\authornote{Mark co-first authors with equal contribution.}
\orcid{0000-0001-8326-1369}
\affiliation{%
  \institution{Simon Fraser University}
  \country{Canada}
}

\author{Yancheng Cao}
\authornotemark[1]
\orcid{0000-0003-3033-8881}
\affiliation{%
  \institution{Columbia University}
  \country{USA}
}

\author{Will Ke Wang}
\orcid{0000-0003-1444-5468}
\affiliation{%
  \institution{Columbia University}
  \country{USA}
}

\author{Yuanzhe Dong}
\orcid{0009-0006-2013-1157}
\affiliation{%
  \institution{Stanford University}
  \country{USA}
}

\author{Changchang Yin}
\orcid{0000-0002-6540-6365}
\author{Weidan Cao} 
\orcid{0000-0001-5417-2121}
\author{Ping Zhang}
\orcid{0000-0002-4601-0779}
\affiliation{%
  \institution{The Ohio State University}
  \country{USA}
}

\author{Jingzhe Yang}
\orcid{0000-0003-4019-0999}
\affiliation{%
  \institution{Nationwide Children's Hospital}
  \country{USA}
}

\author{Bingsheng Yao}
\orcid{0009-0004-8329-4610}
\affiliation{%
  \institution{Northeastern University}
  \country{USA}
}

\author{Yifan Peng}
\orcid{0000-0001-9309-8331}
\affiliation{%
  \institution{Weill Cornell Medicine}
  \country{USA}
}

\author{Chunhua Weng}
\orcid{0000-0002-9624-0214}
\author{Randy Auerbach}
\orcid{0000-0003-2319-4744}
\author{Lena Mamykina}
\orcid{0000-0001-5203-274X}
\affiliation{%
  \institution{Columbia University}
  \country{USA}
}

\author{Dakuo Wang}
\authornote{Mark corresponding authors.}
\orcid{0000-0001-9371-9441}
\affiliation{%
  \institution{Northeastern University}
  \country{USA}
}

\author{Yuntao Wang}
\authornotemark[2]
\orcid{0000-0002-4249-8893}
\affiliation{%
  \institution{University of Washington}
  \country{USA}
}

\author{Xuhai Xu}
\authornotemark[2]
\email{xx2489@columbia.edu}
\orcid{0000-0001-5930-3899}
\affiliation{%
  \institution{Columbia University}
  \country{USA}
}


%
\renewcommand{\shortauthors}{Ying \& Cao et al.}
\renewcommand{\shorttitle}{\projectname{}}

%
\begin{abstract}
While just-in-time interventions (JITIs) have effectively targeted common health behaviors, individuals often have unique needs to intervene in personal undesirable actions that can negatively affect physical, mental, and social well-being.
We present \projectname, a smartwatch-based JITI system that empowers users to define custom interventions for these personal actions with a small number of samples.
For the model to detect new actions based on limited new data samples, we developed a few-shot learning pipeline that finetuned a pre-trained inertial measurement unit (IMU) model on public hand-gesture datasets.
We then designed a data augmentation and synthesis process to train additional classification layers for customization.
Our offline evaluation with 26 participants showed that with three, five, and ten examples, our approach achieved an average accuracy of 76.8\%, 84.7\%, and 87.7\%, and an F1 score of 74.8\%, 84.2\%, and 87.2\%
We then conducted a four-hour intervention study to compare \projectname against a rule-based intervention. Our results demonstrated that our system led to a significant reduction by 64.0$\pm$22.6\% in undesirable actions, substantially outperforming the baseline by 29.0\%.
Our findings underscore the effectiveness of a customizable, AI-driven JITI system for individuals in need of behavioral intervention in personal undesirable actions.
We envision that our work can inspire broader applications of user-defined personalized intervention with advanced AI solutions.

\end{abstract}

%
%
\ccsdesc[500]{Human-centered computing~Ubiquitous and mobile computing}
\ccsdesc[500]{Applied computing~Life and medical sciences}
%
\keywords{Few-shot learning, Just-in-time intervention, Personalized intervention}

%
\maketitle

\begin{figure*}[!t]
\centering 
\includegraphics[width=1\linewidth]{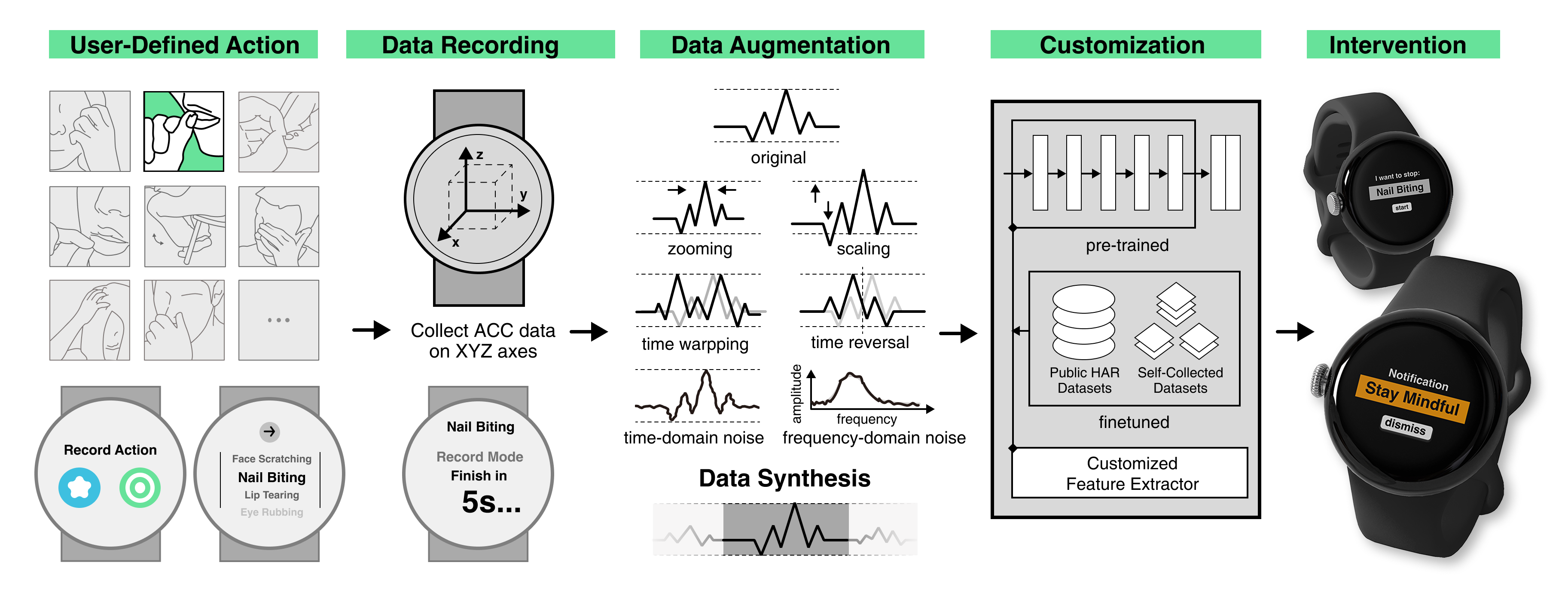}
\caption{\projectname empowers users to easily define personal actions that they want to receive just-in-time intervention (JITI) from a smartwatch. The user journey is as follows: (1) Users determine one or more custom target actions. (2) They follow the instructions on the smartwatch to collect a small set of samples with the accelerometer sensor. (3) \projectname{} applies multiple data augmentation and data synthesis techniques to expand the training dataset, (4) \projectname{} adapts a pre-trained model through fine-tuning and personal customization. (5) \projectname{} leverages the custom model to provide a JITI system for real-time action recognition and intervention delivery.}
\label{fig:teaser}
\Description{}
\end{figure*}

\section{Introduction}
\label{sec:introduction}


Recent advances in mobile sensing technologies and artificial intelligence (AI) have led to the emergence of research on intelligent, just-in-time interventions (JITIs)  using mobile or wearable devices \cite{li2024stayfocused, sarker2014assessing, liao2020personalized, rojas2021scalable, alharbi2023smokemon, rabbi2015mybehavior, zhao2023affective}.
A typical research paradigm usually starts by identifying a target undesirable behavior, followed by data collection from mobile and/or wearable devices, machine learning (ML) model development, and finally, real-time system evaluation (\eg \cite{orzikulova2024time2stop, han2022stressbal, lee2017itchtector, kim2022prediction, alharbi2023smokemon}).
When deployed, these systems will detect the occurrence of target behaviors and deliver JITIs to help users regulate their behaviors and achieve personal health goals.
In the past decade, researchers have achieved a wide range of successful JITI applications, such as reducing smartphone overuse \cite{lu2024interactout, xu2022typeout}, prevention of sedentary habits \cite{liao2020personalized}, smoking cessation \cite{alharbi2023smokemon}, promoting skin health \cite{lee2017itchtector, rojas2021scalable}, and managing stress and emotions \cite{kim2022prediction, koch2021drivers}.

Existing research predominantly focuses on \textit{common} health behaviors that are generally applicable to a large group of populations. However, some individuals' undesirable behaviors can be highly \textit{personal} and \textit{idiosyncratic}. This is especially the case for personal micro-actions or micro-habits.
Example actions include leg-shaking, nail-biting, hair-pulling, and skin-picking (some referred to as body-focused repetitive behaviors, BFRBs) \cite{stein1998phenomenology, snorrason2012skin,teng2002body,oshio2018shake,stein2008trichotillomania}. Such micro-actions can have negative impacts on ones' health (\eg lip-picking can cause cheilitis symptoms \cite{chalkoo2016exfoliative,greenberg2017diseases}), or unfavorable social implications (\eg leg-shaking is considered rude and disrespectful in some cultures~\cite{yilmaz2024leg}).
Such actions vary considerably across individuals \cite{xu2022enabling}, shaped by diverse physical, psychological, social, and environmental factors \cite{trapp2015individual}. Consequently, developing a personalized JITI system poses substantial challenges in both data collection and model training. From a data-collection perspective, it is impractical to require users to gather extensive real-world samples of every undesirable action. From a modeling perspective, training a robust system on such limited data to provide personalized interventions that are adapted to individual contexts and goals remains difficult.

To address this gap, we built \textbf{\projectname}, a smartwatch-based system enabling users to go beyond pre-defined undesirable actions and easily customize new interventions targeted at their own specific undesirable actions.
We developed a few-shot learning pipeline that only requires a small number of samples of the individual target behavior and outputs a reliable ML model for customized behavior detection and real-time JITIs.
Specifically, we built on top of a pre-trained inertial measurement unit (IMU) model based on self-supervised learning (SSL) \cite{yuan2024self} and finetuned the model on multiple open IMU datasets of hand-gesture recognition, with the goal to enhance the model's feature embedding capability on fine-grained actions.
Then, given the small sample sizes of new target behaviors, we adopted a series of data augmentation and data synthesis techniques to train additional lightweight classification layers for the new custom undesirable actions for each individual.

We evaluated our system through both an offline evaluation experiment and a real-time intervention study.
For the offline evaluation, we pre-determined a set of five micro-actions that are typically considered negative behaviors and can be captured with a wrist-worn smartwatch, including face-scratching, nail-biting, eye-rubbing, lip-picking, and leg-shaking \cite{stein1998phenomenology, snorrason2012skin,teng2002body,oshio2018shake,stein2008trichotillomania}.
We then collected data from participants (N=26) on these five actions. We also ask participants to self-define new wrist-based actions that they want to receive intervention. Our final model achieves an average accuracy of 76.8\%, 84.7\%, and 87.7\%, and an F1 score of 74.8\%, 84.2\%, and 87.2\% with one, five, and ten examples.
Building on the model, to evaluate the intervention effectiveness of \projectname, we conducted another four-hour-long study (N=21) that simulated real-life intervention experience.
We compared our system against a rule-based intervention method in an environment where participants naturally tended to perform their self-chosen actions.
The results indicate that \projectname reduced undesirable actions by 64.0±22.6\% with statistical significance (p<0.05), and our system substantially outperformed the baseline intervention method by 29.0\% (p<0.05).
Participants' qualitative feedback also revealed interesting insights into the human-AI intervention experience, including participants' distorted perceptions of the intervention's strength and effectiveness, and various collaborative relationships between users and AI.
The effectiveness of \projectname to mitigate personal undesirable behaviors, shown by both an offline evaluation experiment and a real-time intervention study, sheds light on the future design of personalized AI-powered JITI systems.
Overall, our contributions can be summarized as:

\begin{s_itemize}
\item We introduced \projectname, the first smartwatch-based JITI system that empowers users to define personalized intervention on undesirable micro-actions.
\item We conducted an offline evaluation of our few-shot learning pipeline by recognizing different numbers of undesirable actions and numbers of few-shot samples. This extensive evaluation indicates the robust performance of our pipeline.
\item We implemented \projectname as a real-time intervention system and conducted a user study to evaluate its effectiveness. Our results not only show its advantage over the baseline, but also reveal a range of interesting insights that can guide the future design of human-AI intervention systems.
\end{s_itemize}

\section{Related Work}
\label{sec:background}

In this section, we first provide a general overview of just-in-time behavior intervention, and then a review of prior work in hand gesture recognition based on wearable devices. 

\subsection{Sensing-based Just-in-Time Intervention (JITI)}

Advances in mobile sensing technologies enable the unobtrusive real-time monitoring of individual states and environmental contexts, while delivering proactive cues and user-specific information \cite{choi2019multi}.
Such advances facilitate the implementation of just-in-time intervention (JITI) \cite{nahum2018just}, with the goal of delivering timely and appropriate support for users.  
Early research has applied JITI to address a variety of health-related issues using rule-based approaches \cite{choi2019multi, sun2020beactive, haliburton2023exploring, luo2018time, kim2019lockntype, kim2019goalkeeper, howe2022design, raether2022evaluating, hsu2014persuasive}.
These approaches usually depended on predefined sets of rules and conditions to trigger interventions, which are typically defined by domain experts. Examples include event-based rules \cite{kim2019lockntype, kim2019goalkeeper, raether2022evaluating}, time-based rules \cite{sun2020beactive, haliburton2023exploring, luo2018time}, combinations of multiple rules \cite{howe2022design}, multi-stage rules \cite{choi2019multi}, to name a few. 

Recently, with the advancement of AI techniques, an increasing number of studies have started to apply AI for JITI \cite{orzikulova2024time2stop, wu2024mindshift, xu2022typeout, li2024stayfocused, sarker2014assessing, lee2017itchtector, liao2020personalized, kim2022prediction, rojas2021scalable, alharbi2023smokemon}. In contrast to rule-based JITI, AI-based approaches utilize large-scale user behavior data and trained AI/ML models to determine optimal intervention timing and personalized interventions.
For instance, Time2Stop \cite{orzikulova2024time2stop} employs machine learning to develop an adaptive, explainable intervention system for smartphone overuse that determines optimal timings, offers transparent AI explanations, and integrates user feedback to improve the model over time. Rabbi et al. \cite{rabbi2015mybehavior} and Liao et al. \cite{liao2020personalized} incorporated reinforcement learning algorithms into JITI systems to personalize the model for each user, enhancing the effectiveness of physical activity interventions.


However, these studies primarily focused on predefined ``common'' health behaviors that are broadly applicable to large populations, failing to address personal/idiosyncratic undesirable behaviors that are specific and unique to individual users \cite{stein1998phenomenology, snorrason2012skin,teng2002body,oshio2018shake,stein2008trichotillomania}. 
Idiosyncratic behaviors naturally mean that the sample size (from a single individual) would be much more limited than other common health behaviors, posing challenges to training AI models for intelligent intervention systems.
The limitation of existing solutions reduces the intervention systems' ability to adapt to personal behaviors.
To bridge this gap, our work proposes a personalized intervention approach to deliver customized JITI for user-defined undesirable actions.


\subsection{Wearables for Hand Gesture Recognition and Customization}

The field of hand gesture recognition or activity recognition using wearable technologies has been extensively studied, utilizing a range of sensing techniques (\eg vision \cite{gong_wristwhirl_2016,kim_digits_2012,wu_back-hand-pose_2020,yeo_opisthenar_2019,hu_fingertrak_2020, nguyen2023hand, qi2024computer,xia2024ts2act}, sound wave \cite{nandakumar_fingerio_2016,laput_sensing_2019,harrison_skinput_2010,iravantchi_beamband_2019,iravantchi_interferi_2019, lee2024echowrist, li2023enabling}, electromyography \cite{saponas_enabling_2009,saponas_demonstrating_2008, meng2022user,chamberland2023novel}, pressure or stretch \cite{dementyev_wristflex_2014,jung_wearable_2015,strohmeier_flick_2012, si2022flexible, delpreto2022wearable}, magnetism \cite{chen_finexus_2016,chen_utrack_2013,parizi_auraring_2019,yang_magic_2012,kienzle_lightring_2014,sluyters2022hand,byberi2023glovesense}). 
Among them, motion data (\eg acceleration, angular velocity) collected by IMUs are particularly notable for their effectiveness in capturing dynamic hand gestures ~\cite{laput_viband_2016,wen_serendipity_2016,xu_finger-writing_2015,akl_novel_2011,kim_imu_2019, li2023signring,sharma2023sparseimu}. Coupled with their cost-effectiveness and widespread availability in commercial wearable devices, IMUs' are the best choice of sensor for a JITI system to ensure effectiveness, ubiquity and generalizability.

Existing gesture recognition approaches can be broadly categorized into trajectory-based and ML-based. 
Early trajectory-based methods \cite{liu_uwave_2009, mckenna_comparison_2004} achieve high accuracy with relatively few samples, but they are limited in recognizing more complex gesture trajectories \cite{mckenna_comparison_2004}. 
For more sophisticated and fine-grained gestures, ML-based methods are more suitable, but are typically heavily data-driven, requiring a large number of samples of a pre-defined gesture set to train either a traditional model \cite{georgi_recognizing_2015, iravantchi_beamband_2019,hu2020fine,chen2015utd} or a deep learning model \cite{hu_fingertrak_2020,yeo_opisthenar_2019,li2023signring,leng2024imugpt,shen2024mousering} depending on the dataset size. 

In addition to recognizing gestures from predefined sets, some systems support the customization of user-defined gestures, which enhances memorability~\cite{nacenta_memorability_2013}, interaction efficiency~\cite{ouyang_bootstrapping_2012}, and accessibility for individuals with physical disabilities~\cite{anthony_analyzing_2013}. Most notably, incorporating gesture customization in our system enables users to define their own undesirable gestures for intervention.  
However, to ensure a seamless user experience, the data collection process for new gestures must be efficient and limited in scale (\eg no more than 10 samples). Existing approaches (\eg rule-based methods~\cite{avrahami_guided_2001,doring_gestural_2011} and computational techniques~\cite{lou_personalized_2017,anthony_lightweight_2010, mckenna_comparison_2004,ouyang_bootstrapping_2012}) meet this requirement, but are limited to recognizing hand gestures with significant motion, where IMU signals are distinct, and the recognition task is relatively straightforward. 
To identify more sophisticated and fine-grained gestures or actions, there are two common approaches: 1) collect more data of the new gesture to further train models, or 2) fine-tune a pre-trained base model use a limited amount of new samples. Since the first option would impact user experience in real-life applications, the fine-tuning approach is more appropriate. However, fine-tuning a robust fine-grained gesture recognition model with a small number of samples remains a challenging task \cite{stewart_online_2020,wu_one_2012,rahimian_few-shot_2021, xu2022enabling, rahimian2021fs,zou2024pregesnet}. Most related to our work, Xu et al.~\cite{xu2022enabling} collected extensive gesture data to pre-train a robust model and used few-shot samples to fine-tune new gestures, but this work focuses on short-duration gestures (less than one second), which may limit its potential to support users in defining their own undesirable actions for intervention. Furthermore, the work is closed-sourced, with both its dataset and pre-trained model unavailable for public use, hindering reproducibility and broader applicability.

Building on top of prior work to address the challenges of gesture customization, we developed an open-sourced few-shot learning pipeline using public models and datasets. Our system enables the model to quickly adapt to new target gestures or actions with high accuracy using only a few samples.

\section{\projectname Design}
\label{sec:methods}
We designed a few-shot learning pipeline to enable users to define their own undesirable actions with a small number of examples.
We introduce our technical pipeline (Sec.~\ref{sub:methods:system}), the interface and intervention experience design (Sec.~\ref{sub:methods:interface}), as well as the implementation details of \projectname (Sec.~\ref{sub:methods:implementation}).

\subsection{Few-shot Learning Pipeline}
\label{sub:methods:system}
To achieve the goal of learning user-defined undesirable actions with few-shot samples, we designed and implemented a three-stage pipeline building on public models and datasets.
\autoref{fig:customization_pipeline} visualizes the overall structure of the pipeline.
To ensure compatibility with existing public models and datasets, we used tri-axial accelerometer data from the IMU, sampled at 30 Hz.

\subsubsection{Stage 1: Pre-trained SSL Model}
\label{subsub:methods:system:pretrain}

One of the primary challenges in deep learning for human activity recognition (HAR) is the lack of large labeled datasets \cite{yuan2024self, leng2024imugpt}.
Although there exist multiple public human activity recognition (HAR) datasets with ground truth labels, they often have limited size and different sensing modalities, data sizes, task definitions, and collection protocols (e.g., \cite{hu2020fine,chen2015utd,opportunity_activity_recognition_226,matey2023dataset,ada_alevizaki_2022_7092553,mallol_ragolta_2022_6517688,alexander_holzemann_2024_7654684,herrera2019monitoring,ofli2013berkeley,de2009guide,kirmizis_2021_4507451,kyritsis_2021_4421861,vaizman2017recognizing, hu_2022_7058383}). Therefore, it is challenging to unify these datasets into a single large-scale dataset for pre-training a supervised-learning based model.
SSL addresses these challenges by leveraging vast amounts of unlabeled data to learn meaningful representations through pretext tasks, making it particularly well-suited for HAR tasks \cite{saeed2019multi}.

\begin{figure*}[]
\centering 
\includegraphics[width=\linewidth]{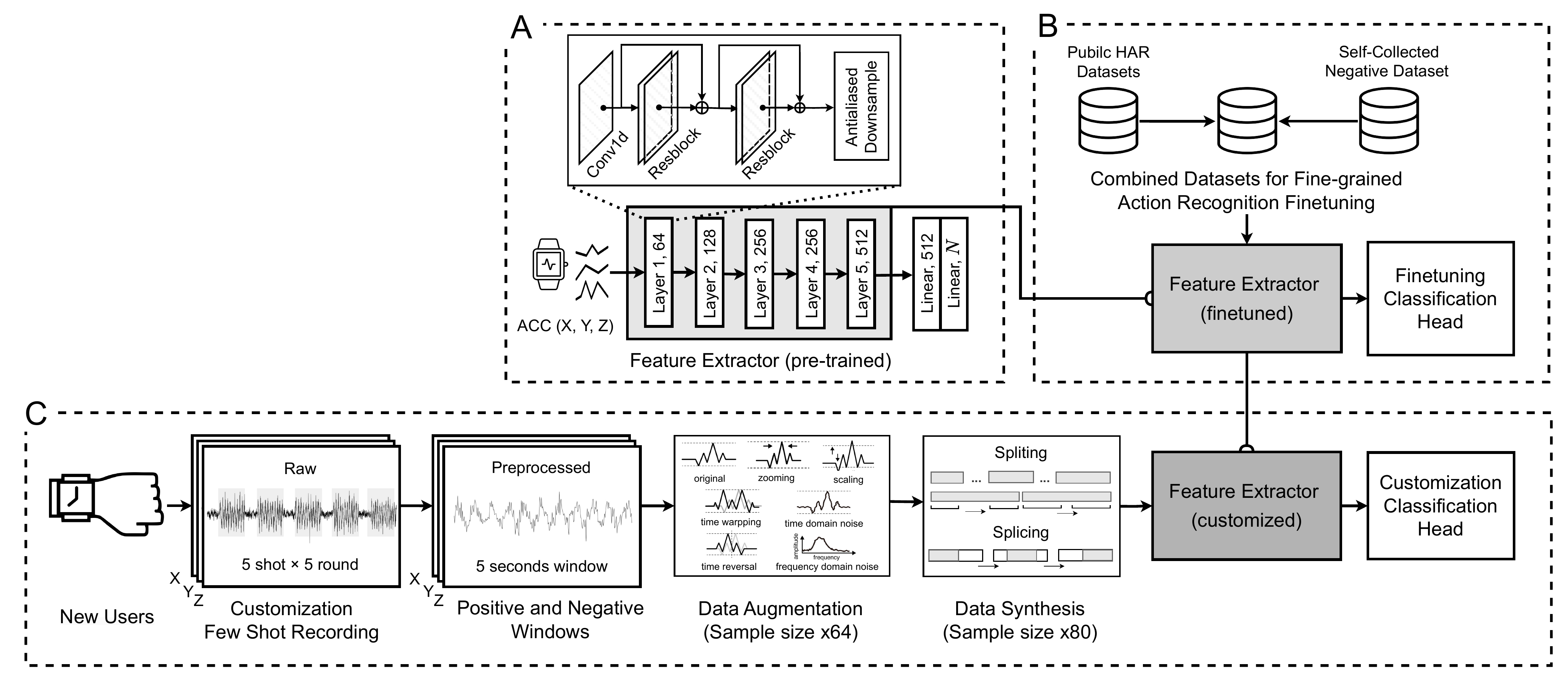}
\caption{Three-stage Few-shot Pipeline for Model Customization.
(A) Stage 1: We adopted A pre-trained SSL model for human activity recognition that takes 30 Hz tri-axis accelerometer data streams.
(B) Stage 2: We finetuned the pre-trained model on two human activity recognition datasets with more fine-grained gestures, together with additional negative data collected by us.
(C) Stage 3: Given the data sequence of a few samples of the new target action, we designed a series of data augmentation and synthesis techniques to enable robust modeling training for customization.
}
\label{fig:customization_pipeline}
\Description{}
\end{figure*}

We adopted a pre-trained model developed by \citet{yuan2024self}, which was trained using multi-task self-supervised learning on the UK Biobank activity tracker dataset \cite{doherty2017large}. The dataset contains over 700,000 person-days of unlabeled wearable sensor data collected from free-living activities via a wrist-worn accelerometer.
This dataset setup fits well into our target application scenarios.
\autoref{fig:customization_pipeline}(A) provides a high-level overview of the architecture of the pre-trained model, which includes a five-layer ResNet-based feature extractor \cite{he2016deep} followed by two fully connected layers.
The model was pre-trained on three fixed-length signals (\ie 5 sec, 10 sec, 30 sec) at 30 Hz.
Three data augmentation techniques \cite{um2017data}, including Arrow of time (AoT), Permutation, and Time warping (TW), were used in pre-training process as three self-supervised tasks \cite{saeed2019multi}.
This pre-trained model is publicly available\footnote{See the model and implementation at: \hyperlink{https://github.com/OxWearables/ssl-wearables}{https://github.com/OxWearables/ssl-wearables}}.

\subsubsection{Stage 2: Model Finetuning}
\label{subsub:methods:system:finetune}

While direct feature extraction from IMU signals using the pre-trained model has demonstrated improved performance in downstream classification tasks~\cite{yuan2024self}, there is a significant gap between the pre-trained tasks (\ie mostly coarse-grained human activities that involves large range of motion) and our target customization tasks (\ie actions with fine-grained activity).
This brings up the need for better model adaptation to bridge this discrepancy.
To address this gap, we finetuned the pre-trained SSL model using two datasets curated by \citet{hu_2022_7058383} and \citet{bhattacharya2022leveraging} that are closer to our use cases in a supervised manner\footnote{The two datasets are available at \hyperlink{https://zenodo.org/records/7058383}{https://zenodo.org/records/7058383} and \hyperlink{https://doi.org/10.18738/T8/NNDFQD}{https://doi.org/10.18738/T8/NNDFQD}}.
These datasets contain labeled, fine-grained, hand-specific human activity data.

We pre-processed the datasets before merging them together. We first unified their units, resampled them to 30 Hz to maintain uniformity across datasets, and then applied z-score normalization to standardize the signal. After these steps, we adopted a 5-second sliding window with a step size of 0.1 second, because a 5-second window fits one of the pre-training signal lengths in the SSL model in Sec.~\ref{subsub:methods:system:pretrain} and is appropriate for our use cases (the same window size as described in Sec.~\ref{subsub:methods:system:fewshot}).
As for the labels, we manually unified activities with the same semantics (\eg handwriting in \cite{hu_2022_7058383} and writing in \cite{bhattacharya2022leveraging}), resulting in a combined dataset with 26 activity classes plus one negative class (no target action).

To establish a more comprehensive and diverse representation of the negative class, we recruited a small group of participants (N=10), each performing 30 minutes of regular indoor daily activities, such as sleeping, playing computer games, studying, and cooking. Participants were specifically instructed and supervised to minimize potentially undesirable micro-actions to ensure the quality of negative data. We manually remove the improper data episodes from the data. This data was then sampled for the subsequence training process. This step enables the model to learn from more comprehensive and diverse negative class samples, thereby reducing false positives.

Combining all these datasets together resulted in a dataset with about 12.5~hours of positive signals and 5 hours of negative signals (before sampling) in total. Using this combined dataset, we performed finetuning on the pre-trained model, with all layers activated (see \autoref{fig:customization_pipeline}(B)).
We adopted weighted cross-entropy as the loss function to address the class imbalance problem.

\subsubsection{Stage 3: Few-Shot Model Customization}  
\label{subsub:methods:system:fewshot}

As mentioned earlier, in real-world applications, it is often impractical for users to provide a large number of samples of a self-defined action for model training.
To achieve individual customization, our few-shot learning procedure was designed to train a new prediction head with lightweight layers built upon the finetuned model from Sec.~\ref{subsub:methods:system:finetune}.
For easy understanding, we will explain our pipeline with the case of adding one intervention action (\ie binary classification) in detail, starting with data collection and then the few-shot learning process. The scenarios with multiple actions adopt the same method.

We implemented a simple, user-friendly data collection process for customization, where a user would follow instructions on the wearable device to repeat the target action several times (N shots, 10 seconds each time), with a short period of 5-second pause or other activities (negative class) between the two repetitions, as shown in \autoref{fig:customization_pipeline}(C).
Given such a signal sequence, we applied the same sliding window process as in Sec.~\ref{subsub:methods:system:finetune} (5-second width and 0.1-second step).
Each window was labeled as positive if it contained more than 3 seconds of the target action; otherwise, it was annotated as negative.

We then introduced a signal processing procedure, including both data augmentation and data synthesis, to enhance our data for model training.
First, we adopted six data augmentation techniques~\cite{iglesias2023data}:
1) zooming, to simulate variations in action speed, randomly selected from $\times0.9$ to $\times1$;
2) scaling, to represent variations in action intensity, with the scaling factor $s \sim \mathcal{N}(1,0.2^2), s \in [0,2]$;
3) time warping, to simulate action temporal variance, using 2 interpolation knots and a warping randomness $w \sim \mathcal{N}(1,0.05^2), w \in [0,2]$;
4) time reversal, to simulate temporal variation, by reversing the action's time sequence;
5) time-domain noise, to simulate sensor inaccuracies or environmental disturbances, Gaussian noise with a noise level of 0.01 is added to the original data;
and 6) frequency-domain noise, to simulate frequency variation or external interferences. We add the random noise to the frequency components of the signal (after Fourier transform) and then transformed the signal back to the time domain (with inverse Fourier transform).
We went through all combinations of these six augmentation technique steps, which increased the size of the data by $2^6-1$ times.

Next, we designed a data synthesis step. To create additional samples that simulate short episodes of target undesirable actions, we artificially concatenated short segments of the target undesirable action with negative episodes (no target undesirable actions).
The positive segments (of the target undesirable actions) were chosen randomly from the 10-second continuous recordings and would have varying lengths sampled from [3, 4.9] seconds.  
The starting position of the positive episode was also randomized within a 5-second window, with negative episodes padding at the beginning and the end.
This step further increased the sample size by about 80 times.
In total, our data augmentation and data synthesis steps enlarged the original few-shot samples by about 143 times.

Finally, to customize the model to recognize a new target undesirable action, we further trained the finetuned model with a new classification head with the total set of data. The head was trained for a binary classification when adding one intervention action, and N+1-class classification when adding N new actions.

In the real-time system, the final classification model followed the same sliding window setup and performed classification at 10 Hz (0.1-second step).
To improve the system robustness, we added a smoothing step with the threshold as 3 based on grid search, \ie the system will only recognize a target action if there is a consecutive sequence of positive outcomes from 3 windows.

\subsection{Intervention Design}
\label{sub:methods:interface}
\begin{figure*}[t]
\centering 
\begin{subfigure}[t]{0.48\textwidth}
    \centering
    \includegraphics[height=5.2cm]{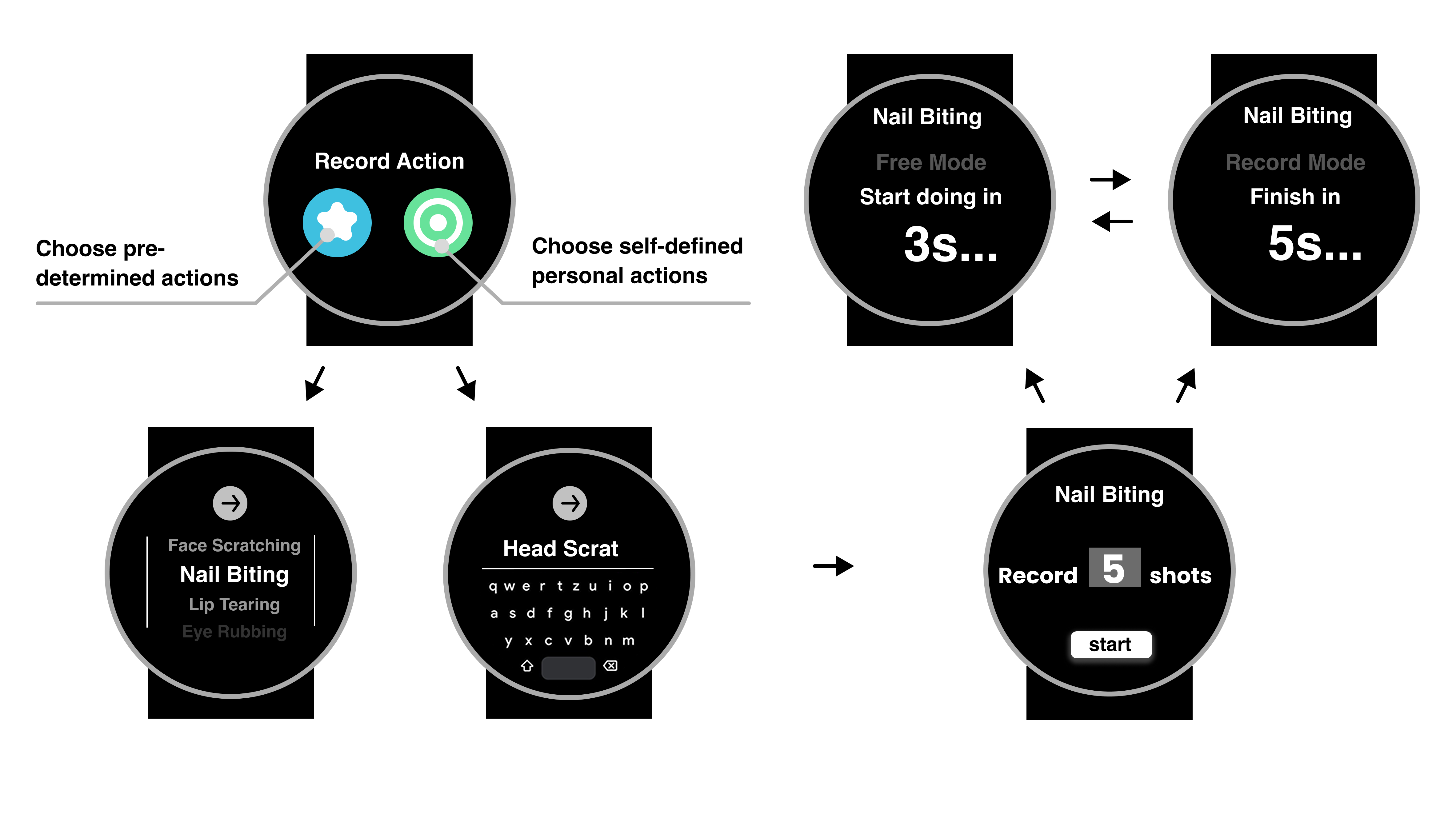}
    \caption{Few-shot Data Collection Interface}
    \label{subfig:interface_design:data_collection}
\end{subfigure}
\hfill
\begin{subfigure}[t]{0.48\textwidth}
    \centering
    \includegraphics[height=5.2cm]{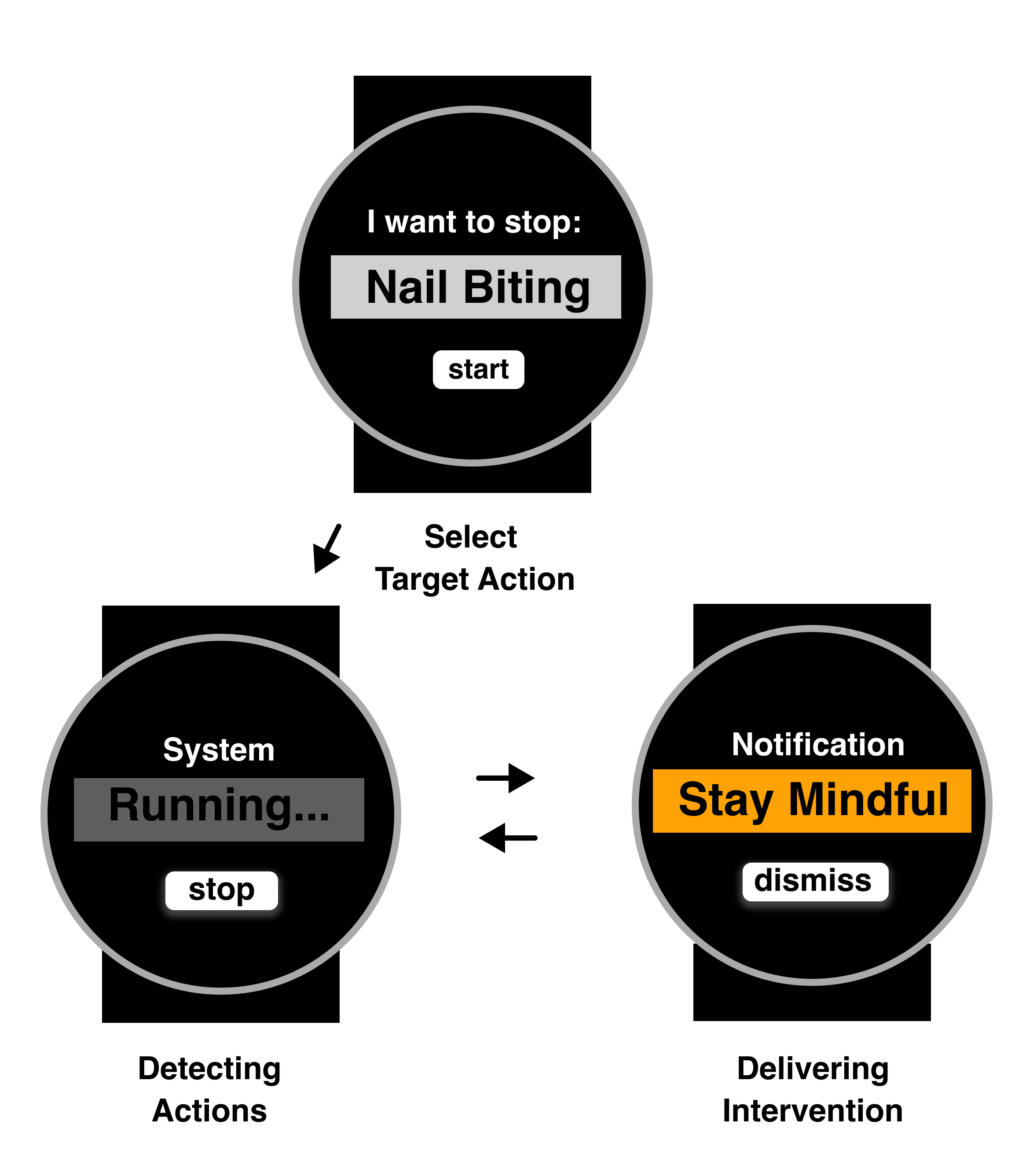}
    \caption{Intervention Reminder Interface}
    \label{subfig:interface_design:intervention}
\end{subfigure}
\caption{Smartwatch Interface Designs. (a) Few-shot data collection interface, where a user can define the target behavior and the number of shots. The user can name the gesture once the collection is finished. (b) Intervention reminder interface, which is shown when the system detects undesirable target actions.}
\label{fig:interface_design}
\Description{}
\end{figure*}

Building upon the customized model, we then developed a real-time intervention system on the smartwatch.
As introduced in Sec.~\ref{subsub:methods:system:fewshot}, in the initial customization process, a user would go through a simple data recording to collect a few samples of the undesirable action.
\autoref{subfig:interface_design:data_collection} shows the interface on the smartwatch to select or define an undesirable action and set up the number of shots for customization.

Once a model was trained following the 3-stage pipeline, the user could enter the live-stream mode to receive interventions.
Whenever the target action was detected, the watch would send a reminder notification with vibration. \autoref{subfig:interface_design:intervention} shows the interface of the intervention.
The user could click a button to dismiss the reminder.
To avoid delivering overwhelming interventions, we set a 5-minute cool-down time, \ie at most, one intervention can be delivered during this interval.

\subsection{System Implementation}
\label{sub:methods:implementation}

We adopted a client-server architecture for the system implementation to enable efficient data transmission and processing. The interface was implemented on the Google Pixel Watch 2, which acted as the client device. It continuously streamed the accelerometer data, collected at 30 Hz, to a dedicated server in real time via a socket communication protocol. 

Before real-time data transmission, the server had already completed the initial setup stages, namely Stage 1 and Stage 2, as described in Sec.~\ref{sub:methods:system}. Therefore, once the customization data from the client was collected, we utilized an A100 GPU to perform the few-shot custom model training, enabling rapid adaptation to new data with minimal samples. 

After training the model on the server, we deployed the final model for real-time inference. The inference process ran on the server, and the results were transmitted back to the client for immediate feedback, enabling efficient and responsive action recognition and the delivery of JITI.

\section{Model Evaluation}
\label{sec:model_evaluation}
In this section, we report the evaluation of \projectname's few-shot learning pipeline offline performance. We will further elaborate on the evaluation of \projectname's intervention effectiveness in section \ref{sec:intervention_evaluation}.

\subsection{Data Collection}
\label{sub:model_evaluation:data_collection}
\subsubsection{Participants}
We recruited 26 users (14 females, 12 males, age 22$\pm$2) for data collection via social media platforms.
We focused on users who were aware of their own undesirable actions and had the intention to reduce these actions. These are the target users of our intervention system.
Our study was IRB-approved by the local institution, and participants were compensated with \$10 for this data collection study (around 45 minutes).

\subsubsection{Personal Undesirable Action Customization}

\begin{figure*}[]
\centering 
\includegraphics[width=1\linewidth]{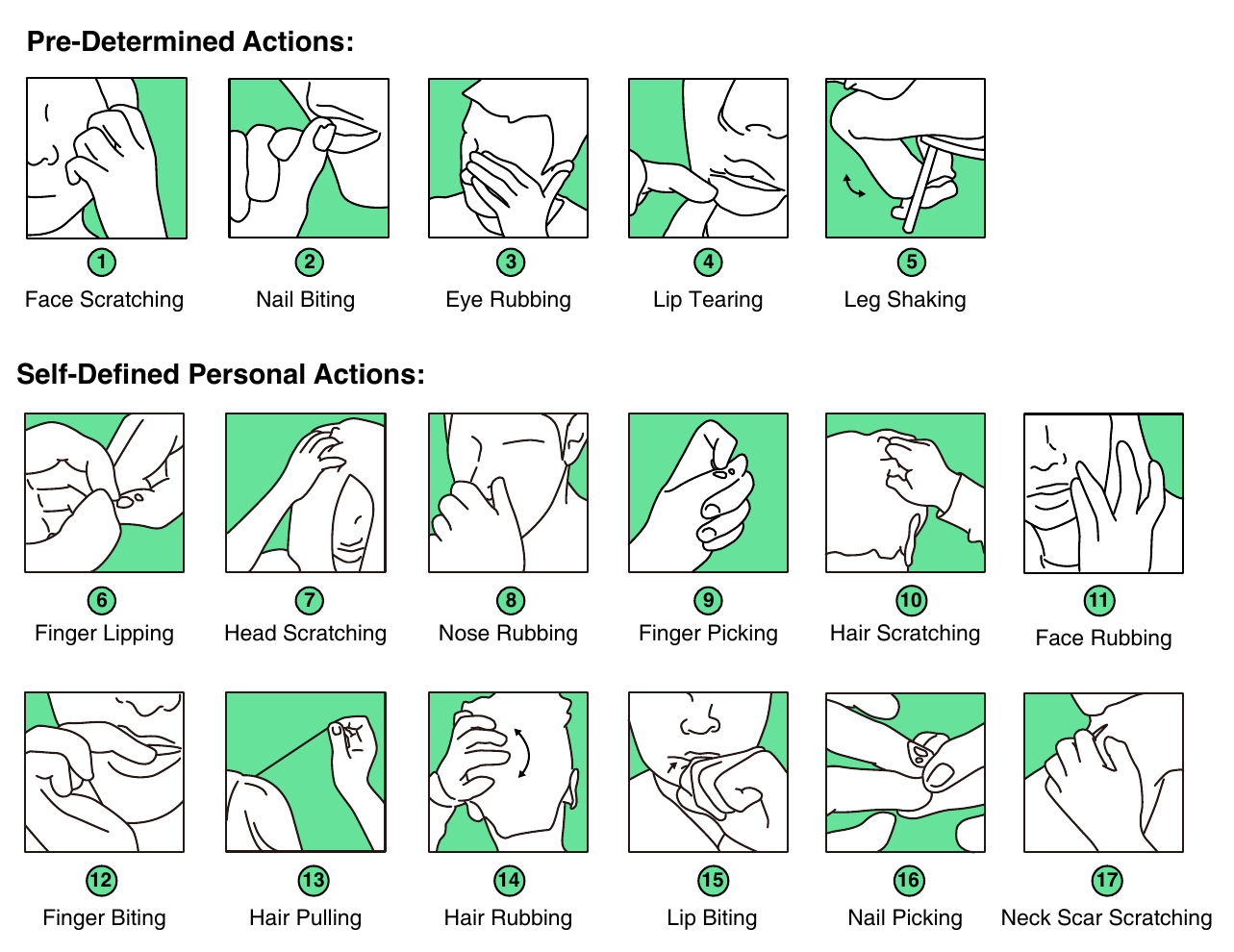}
\caption{Target Actions for Evaluation. (1-5) presents the five pre-determined actions. (6-17) visualizes new target behaviors defined by participants.
Only identical actions are grouped as one. Actions that have minor differences are counted separately, as each of them could be highly personal.
}
\label{fig:evaluation_actions}
\Description{}
\end{figure*}

Participants were asked to record five pre-determined target actions that are commonly recognized as undesirable actions~\cite{teng2002body,oshio2018shake}, including \textit{Face Scratching}, \textit{Nail Biting}, \textit{Eye Rubbing}, \textit{Lip Tearing}, and \textit{Leg Shaking}. The first five figures in \autoref{fig:evaluation_actions} illustrate these actions.

Moreover, each participant was asked to define a new undesirable action tailored to their own personal needs.
In total, 26 participants designed an additional set of 12 actions, including \textit{Finger Lipping} (designed by N=5 participants), \textit{Head Scratching} (N=5), \textit{Nose Rubbing} (N=4), \textit{Finger Picking} (N=3), \textit{Hair Scratching} (N=2), \textit{Face Rubbing} (N=1), \textit{Finger Biting} (N=1), \textit{Hair Pulling} (N=1), \textit{Hair Rubbing} (N=1), \textit{Lip Biting} (N=1), \textit{Nail Picking} (N=1), and \textit{Neck Scar Scratching} (N=1).
We only grouped identical actions and distinguished actions as long as they differed slightly. For instance, \textit{Head Scratching} and  \textit{Hair Scratching} were similar, but one involved contacts between fingers and 
scalp, while the other one did not.
Similarly, \textit{Finger Picking} and \textit{Nail Picking} were also quite close, yet one solely focused on the skin on the finger, while the other focused on nails.
These actions were visualized in the second half of \autoref{fig:evaluation_actions}.

\subsubsection{Data Collection Procedure}
For each action, participants followed a consistent protocol (briefly mentioned in Sec.~\ref{subsub:methods:system:fewshot}) comprising two phases per shot: a 5-second \textit{free mode} and a 10-second \textit{record mode}. In the free mode, participants were free to rest or perform natural daily activities (negative data). Once entering the record mode, they performed the target actions (positive data).
This process was repeated across five rounds, with each round consisting of five consecutive shots.
Participants took a short break between two rounds to prevent physical fatigue and were asked to freely adjust the watch position between each round to increase data variance.
In total, we collected 25 shots for each target action.
Moreover, we leveraged the onboarding process at the beginning of the data collection to passively record participants' natural activities (about 5 minutes). This was used as additional data to augment the negative class\footnote{In real-world applications, we envision that such negative data can also be passively collected and implicitly embedded in the instruction process, thereby introducing minimal additional workload for the user}.

The \textit{free mode} segment was labeled as negative data, while the \textit{record mode} segment was labeled as positive data.
To prevent data contamination, the first two seconds during the record mode were excluded from training because these recordings were mixed with postural changes and arm movement.

\subsection{Offline Performance Evaluation}
\label{sub:model_evaluation:pipeline_evaluation}
We evaluated our pipeline by adding one or more actions as target actions.
For each action, we randomly selected two rounds of recordings as the training set (up to 10 shots), one round as the validation set (5 shots), and the remaining two rounds as the test set (10 shots). We repeated the training three times and calculated the average performance.

It is noteworthy that the model performance has two aspects: the window level and the action level. 
For the window level, each sliding window is counted as a binary classification data point (same as the model training process).
For the action level, windows are aggregated with a smoothing threshold of 3 (Sec.~\ref{subsub:methods:system:fewshot}) and represent a closer experience as real-life applications. Such aggregation significantly reduces the false negative and false positive.

\subsubsection{Prediction Performance with Different Number of Shots and Actions}
\label{subsub:model_evaluation:pipeline_evaluation:shots and action}
We evaluated the model performance by training on one to ten shots of the data.
For action recognition, we started by adding one action for each participant (\ie training binary classification models).
To evaluate the performance of multi-class classification models, we also experimented with customizing multiple actions (up to six, as each participant recorded five pre-designed actions and one custom action). This led to a total number of 63 combinations from one to six actions (\(\sum_{k=1}^6 \binom{6}{k}\)).
In total, we trained and evaluated 49,140 models = 10 shot numbers $\times$ 63 action combinations $\times$ 26 participants $\times$ 3 repetitions. 

\begin{figure*}[]
\centering 
\includegraphics[width=\linewidth]{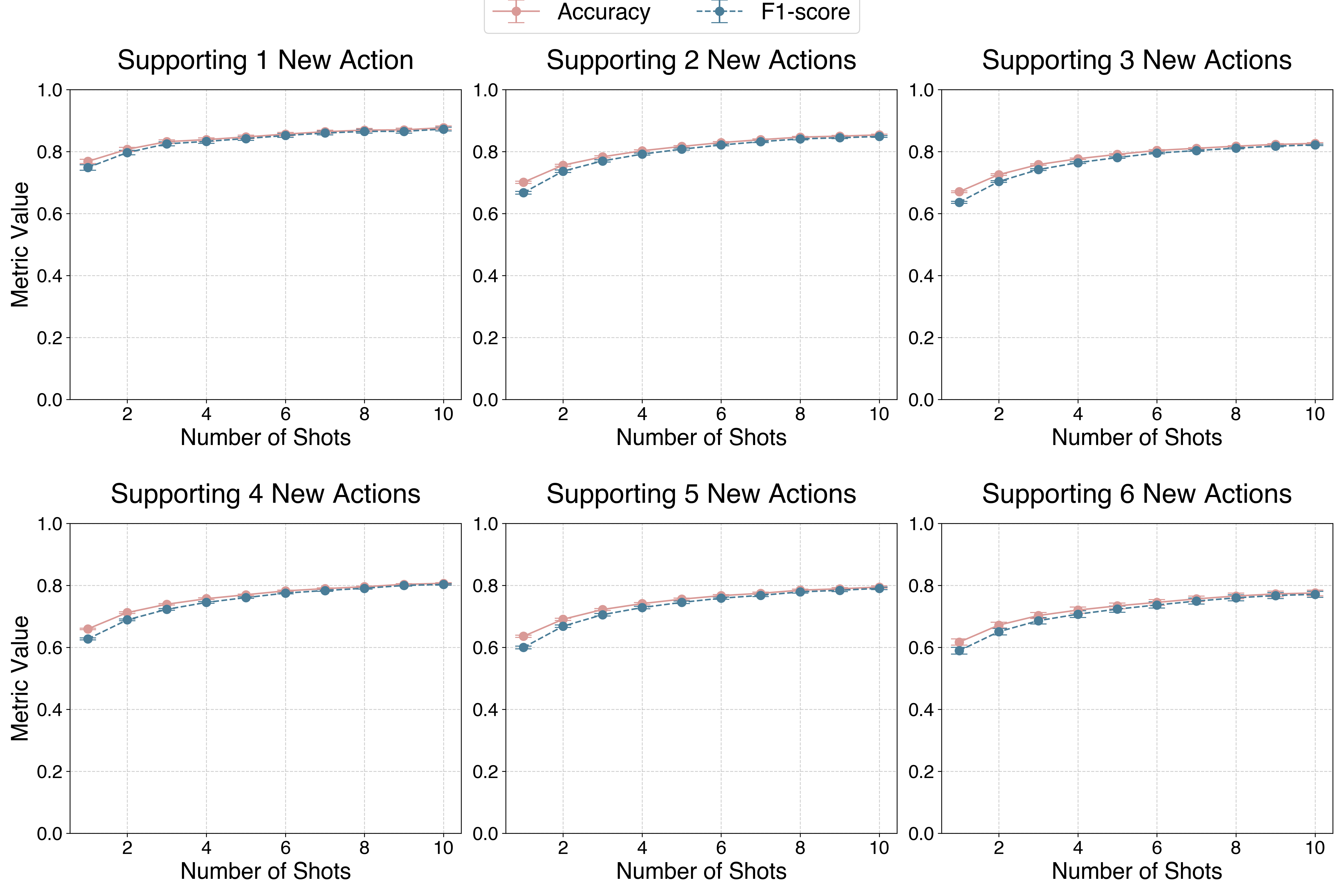}
\caption{Few-shot Learning Pipeline Performance of Accuracy and F1 Score. We experimented with different numbers of shots using 1 to 10 samples to train a custom model. We also experimented with adding more than one target action simultaneously (\ie multi-class classification). Error bars indicate standard error. The same below.
}
\label{fig:all_result_ges_num_action}
\Description{}
\end{figure*}

We mainly focused on the action-level performance. \autoref{tab:action_shot_study} presents both the window-level and action-level results.
As shown in \autoref{fig:all_result_ges_num_action}, when using only one shot to add a new action (\ie the user performs the action only once), our framework achieved an average accuracy of 76.8\% and an F1 score of 74.8\%.
The recognition performance became better with more shots for training the model. With five shots of a new action, our framework attained an average accuracy of 84.7\% and an F1 score of 84.2\%. When using ten shots, our model's performance achieved 87.7\% and 87.2\%, respectively.

Recognizing multiple new actions simultaneously presented a greater challenge. However, compared to the performance of adding one action with five shots (84.7\% and 84.2\%), introducing three new actions (\ie four-class classification) with five shots each, the framework maintained a good average accuracy of 79.1\% and an F1 score of 78.1\%.
Even with six additional new actions and five shots each, the framework still achieved an average accuracy of 73.7\% and an F1 score of 72.3\%.
These results demonstrated the robustness and effectiveness of our pipeline for data-efficient action recognition.

\renewcommand{\arraystretch}{1.3}
\begin{table}[]
\centering
\caption{Detailed Few-shot Pipeline Performance with Different Numbers of Shots when Adding Personal Action.
Window-level results are based on each sliding window as a data point.
Action-level results are the aggregation of the sliding windows after smoothing post-processing (threshold=3) and are closer to real-life application scenarios.
}
\label{tab:action_shot_study}
\resizebox{1\textwidth}{!}{
\begin{tabular}[t]{l|cccc|cccc}
\toprule
\multirow{2}{*}{\textbf{Shots}}  & \multicolumn{4}{c}{\textbf{Window-level}}      & \multicolumn{4}{c}{\textbf{Action-level}}       \\ \cline{2-9}
 & \textbf{Acc}   & \textbf{Prec}  & \textbf{Rec}   & \textbf{F1}    & \textbf{Acc}   & \textbf{Prec}  & \textbf{Rec}   & \textbf{F1}      \\ \hline
1    & 0.614$\pm$0.006 & 0.700$\pm$0.007 & 0.614$\pm$0.006 & 0.571$\pm$0.008 & 0.768$\pm$0.007 & 0.810$\pm$0.007 & 0.768$\pm$0.007 & 0.748$\pm$0.009 \\
3    & 0.658$\pm$0.005 & 0.736$\pm$0.005 & 0.658$\pm$0.005 & 0.634$\pm$0.006 & 0.832$\pm$0.006 & 0.860$\pm$0.005 & 0.832$\pm$0.006 & 0.825$\pm$0.006 \\
5    & 0.670$\pm$0.005 & 0.746$\pm$0.005 & 0.670$\pm$0.005 & 0.648$\pm$0.006 & 0.847$\pm$0.005 & 0.871$\pm$0.005 & 0.847$\pm$0.005 & 0.842$\pm$0.006 \\
7    & 0.685$\pm$0.005 & 0.755$\pm$0.005 & 0.685$\pm$0.005 & 0.667$\pm$0.006 & 0.864$\pm$0.005 & 0.883$\pm$0.005 & 0.864$\pm$0.005 & 0.860$\pm$0.006 \\
10   & 0.702$\pm$0.005 & 0.763$\pm$0.005 & 0.702$\pm$0.005 & 0.688$\pm$0.006 & 0.877$\pm$0.005 & 0.890$\pm$0.005 & 0.877$\pm$0.005 & 0.873$\pm$0.006 \\
\bottomrule
\end{tabular}
}
\end{table}
\renewcommand{\arraystretch}{1.0}

\subsubsection{Prediction Performance of Each New Gesture with Different Number of Shots}
\label{subsub:model_evaluation:pipeline_evaluation:one new gesture}

\begin{figure*}[]
\centering 
\includegraphics[width=\linewidth]{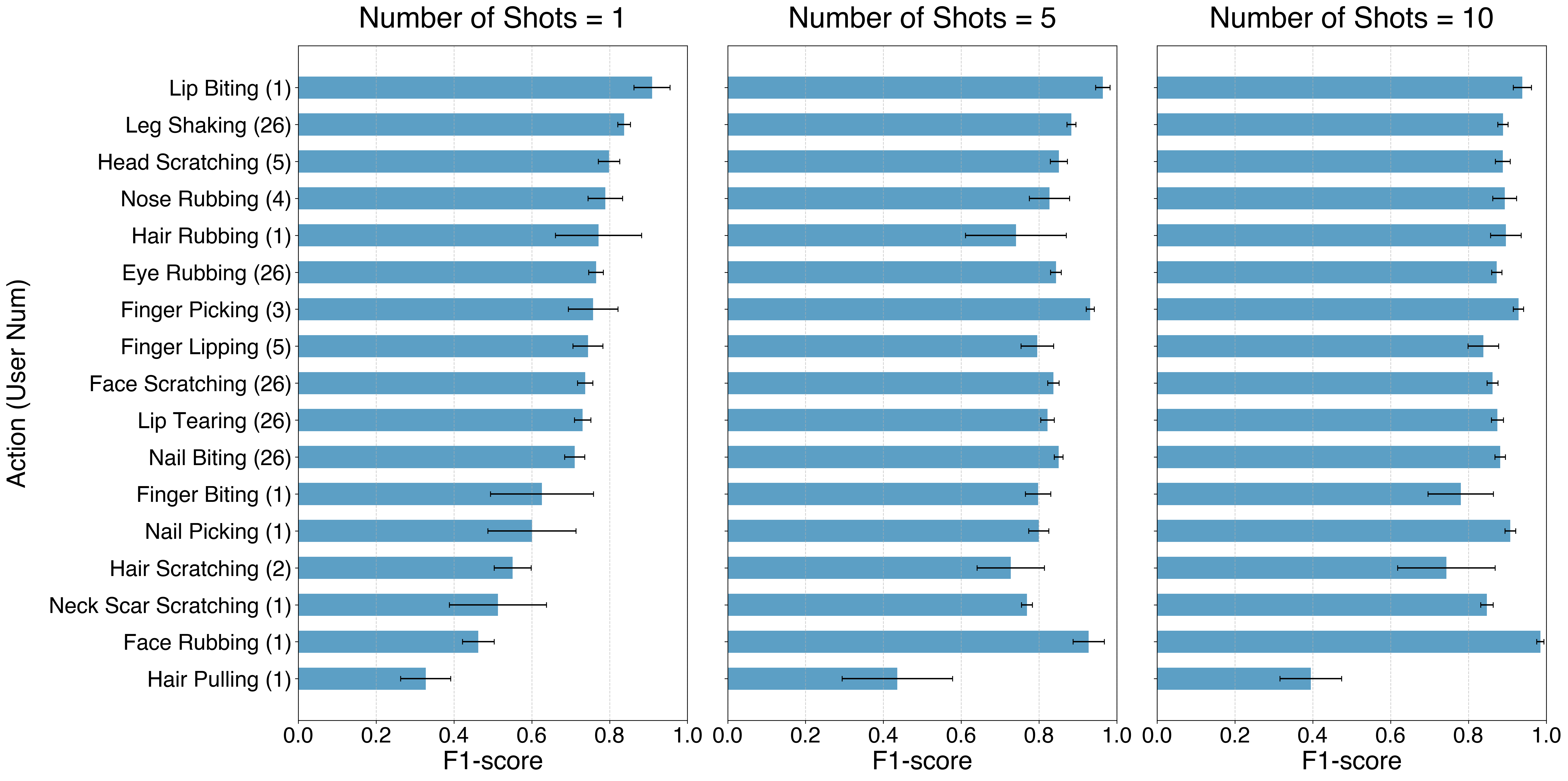}
\caption{Model Performance of Recognizing Each Action with 1, 5, or 10 Shots. For consistency, each action was added alone (\ie binary classification model). The ``(User Num)'' indicates how many users did this action. The five pre-determined actions (Lip Tearing, Nail Biting, Face Scratching, Eye Rubbing, and Leg Shaking) have the total number of participants (26), and other self-defined actions are more scattered.}
\label{fig:all_result_shot_num_action}
\Description{}
\end{figure*}

We further compared the recognition performance across actions.
As shown in \autoref{fig:all_result_shot_num_action}, most of the 17 actions exhibited good performance. Using only one shot, about half of the actions achieved an F1 score above 75\%. When the number of shots increased to five, 14 out of 17 actions surpassed this threshold. With ten shots, performance improved further for most actions, with 12 out of 17 actions achieving an F1 score above 85\%.
\textit{Hair Pulling} appeared to be an exception. Its performance did not improve with more samples after five shots. This was probably due to the overly large variance of the \textit{Hair Pulling} action, even performed by the same individual, and it was challenging for a model to achieve reliable performance even with a limited amount of additional data.

Overall, these results indicate our framework has good learning ability for new actions.

\section{Intervention Evaluation}
\label{sec:intervention_evaluation}
The promising model performance in Sec.~\ref{sub:model_evaluation:pipeline_evaluation} has validated the effectiveness of our few-shot learning pipeline.
Building upon the pipeline, we further conducted a user study to evaluate the effectiveness of \projectname and compared it against a rule-based baseline intervention system.

\subsection{Participants}
\label{sub:intervention_evaluation:participants}
With IRB approval, we recruited the same set of participants in Sec.~\ref{sub:model_evaluation:data_collection} for a follow-up intervention study. 
In the previous data collection, participants performed five per-determined actions and a self-defined action. In this study, they were asked to select one of the six actions that they had the strongest need for intervention.
This action was set as the target action for intervention during the study.
Among the 26 participants, 5 of them did not follow the study protocol. Their results were removed as outliers.  This section focused on the findings based on the remaining 21 participants.

\subsection{Intervention Setting}
\label{sub:intervention_evaluation:setting}
Since personal undesirable actions are inherently difficult to predict or control, we designed an intervention experience that closely mirrors real-life contexts to enhance ecological validity, encouraging participants to perform these actions under more natural conditions.
Our initial conversation with participants indicated two common scenarios where they tended to perform these actions: when they were in an engaging task with a relaxing state (\eg watching an interesting movie or a reality show with dramatic twists and turns); and when they were bored or disengaged (\eg mindlessly scrolling through social media or watching a tedious video) \footnote{Several participants also mentioned the scenarios under pressure or stress. Considering the feasibility and ethics of a multi-hour intervention study, we did not provide this option.}.
Therefore, we set up two types of video-watching tasks and allowed participants to pick the type in which they tended to perform more undesirable actions.

\begin{figure*}[]
\centering 
\begin{subfigure}[t]{1\textwidth}
    \centering
    \includegraphics[width=\textwidth]{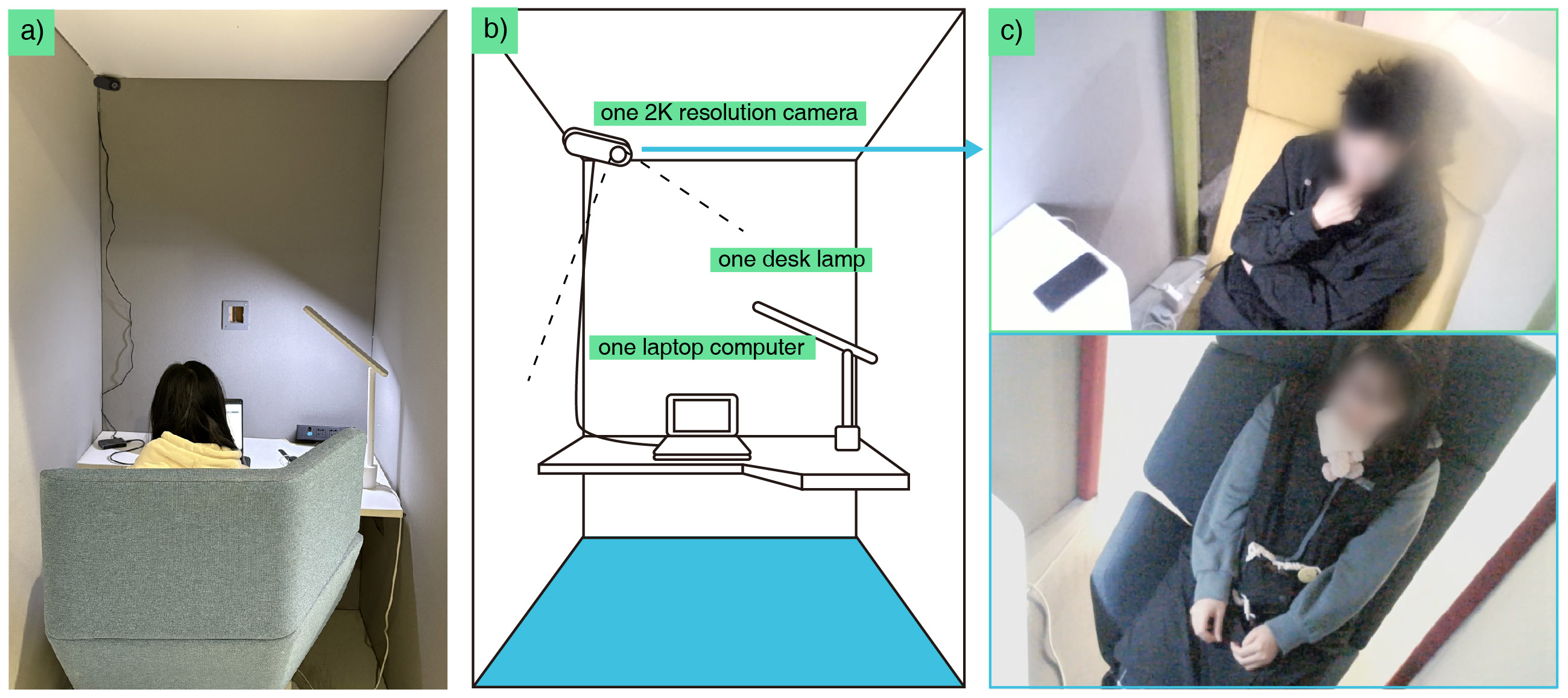}
    \caption{Intervention Room Setup}
    \label{subfig:intervention_setup:design}
\end{subfigure}
\caption{\projectname Intervention Evaluation Setup. (a) The sketch of the study room and apparatus setup for intervention. (b) The photo of a participant in the room. (c) The video from the camera on the corner that records the ground truth.}
\label{fig:intervention_setup}
\Description{}
\end{figure*}

The first type included \textit{engaging} videos. We prepare a set of multi-hour videos for participants to choose from, such as the Harry Potter movie series, sports competitions, and mystery/detective shows.
The second type was watching \textit{disengaging} videos. Examples include cycling or driving route videos, math problem explanations, and public health lecture videos.
Participants sat in a quiet room with a laptop on the table and watched the video they selected, as shown in \autoref{fig:intervention_setup}(a) and (b).
During the video-watching, participants were not interrupted by the experimenter, simulating the real-life setting. 

\subsection{Study Design and Procedure}
\label{sub:intervention_evaluation:design}
We adopted a within-subject design and compared our AI-powered \projectname against a rule-based intervention system.
In the rule-based system, a regular notification (the same interface as \autoref{subfig:interface_design:intervention}) was delivered every 10 minutes, regardless of whether the user did the action.
To mitigate the effect of the two systems outputting different numbers of notifications, we further added restrictions in \projectname so that the number of delivered notifications would be in the range of $\times 0.5$ to $\times 2$ as the baseline system.
This was achieved by forcefully delivering a notification if there was no intervention by the end of each 20-minute window ($\times 0.5$ times of interventions in minimum).
With the 5-min cool-down setup, \projectname can only deliver up to one intervention every 5 minutes, which would be no more than $\times 2$ times of interventions as the baseline.

Our study procedure was designed as follows. After selecting the personal target undesirable action and the task type (engaging vs disengaging), participants would calibrate and familiarize themselves with the intervention system and study setup. They then attended two intervention sessions in total, one session per day. We counterbalanced the order between \projectname and the baseline system, and participants were blind to the order of the two systems.
After familiarizing themselves with the room environment and setup, participants went through each intervention session with three stages (in total 130 minutes): (1) a 30-minute \textit{pre-intervention stage}, where there was no intervention delivered; (2) a 90-minute \textit{intervention stage}, where \projectname or the baseline system would deliver interventions as designed; and (3) a 10-minute \textit{post-intervention} stage, where no more intervention was delivered to observe any lasting effect \footnote{Due to the restrictions of the room booking time and device battery, we regretfully could not do a longer post-intervention stage. We recognize this as a limitation of our study in discussion.}.

The whole intervention session was video-recorded by a camera from the ceiling, positioned at an angle to capture participants' micro-actions and collect ground truth (see \autoref{fig:intervention_setup}(c)). We manually annotated the video and calculated the number and duration of the target actions during the three stages.
We collected participants' Self-Report of Habit Strength of the target action~\cite{verplanken2003reflections} before and after each session.
After the post-intervention stage, we further collected quantitative data from participants with a questionnaire that includes System Usability Scale (SUS) survey~\cite{bangor2008empirical} and Working Alliance Inventory (WAI, short revision)~\cite{munder2010working}.
In addition, we conducted a brief semi-structured interview to collect qualitative feedback on the intervention experience from each participant.

In total, the two sessions took around 5 hours for each participant. To reduce user fatigue, the two sessions were scheduled on two different days within a week. Participants were compensated with \$50 for the intervention study.

\subsection{Intervention Results}
\label{sub:intervention_evaluation:intervention_results}
We first summarize the quantitative results from our study.
We coded the recorded videos by documenting the duration of target actions performed by participants every 10 minutes across the three stages.
Since participants had diverse behavior patterns, we normalized the results with each individual's target action duration in the pre-intervention stage as the reference.
The \textit{relative duration} was calculated by dividing the average duration of target actions per 10 minutes in both the intervention and post-intervention stages by that of the pre-intervention stage. A lower relative duration means more reduction of the target actions compared to the pre-intervention stage.



\begin{figure*}[]
\centering 
\includegraphics[width=\linewidth]{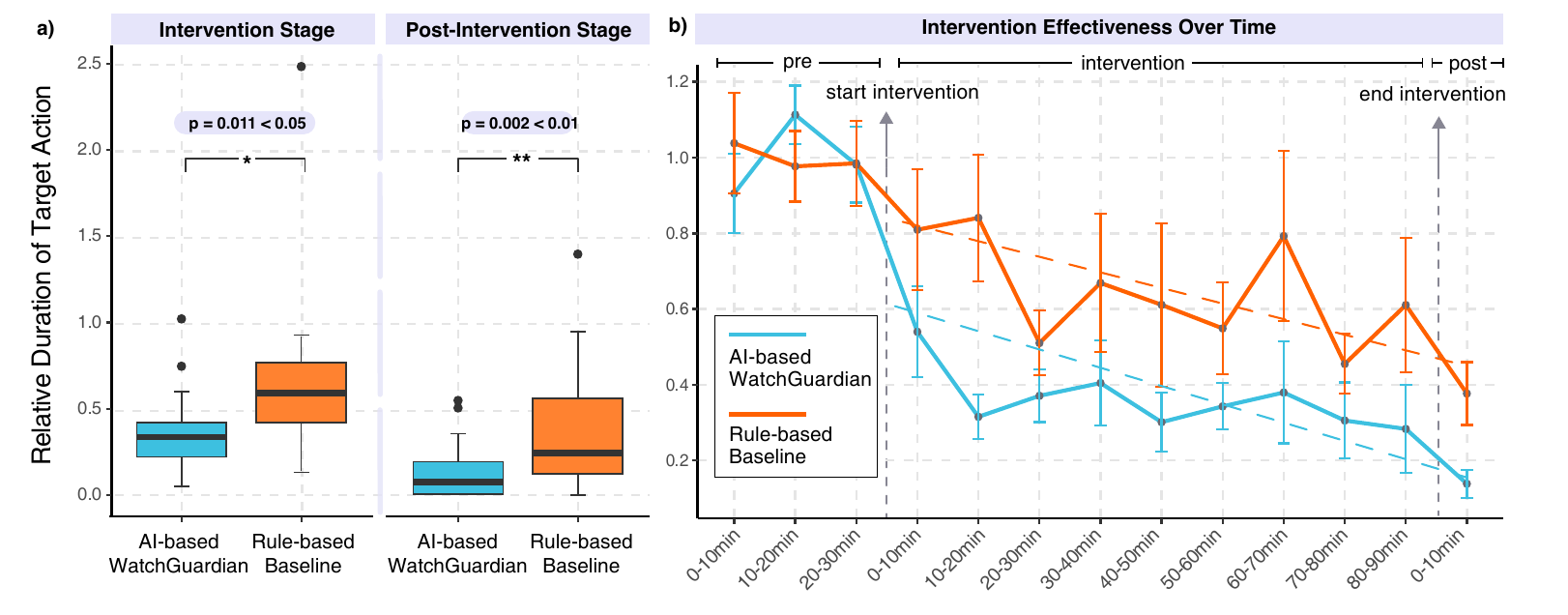}
\caption{(a) Relative Duration of target action every 10 minutes in intervention and post-intervention stages (compared to the pre-intervention stage). A number lower than 1.0 means that an individual performed fewer target actions after intervention.
(b) Average Relative Duration of target action over time. The dashed lines fit the last 10 minutes of the pre-intervention stage and the rest of the session.
}
\label{fig:duration_result}
\Description{}
\end{figure*}

\subsubsection{Reduction of the Duration of Target Actions by Intervention.}
We compare the relative duration between \projectname and the baseline in both intervention and post-intervention stages.
Since participants received slightly more notifications in \projectname during the intervention stage (on average 11.8 \vs 9.0 times per session), we controlled the effect of the number of notifications by using generalized linear mixed models (GLMMs).
Specifically, a GLMM had relative duration as the dependent variable, with the intervention method (AI-based in \projectname \vs rule-based in baseline) and the number of notifications as the main factors.

As shown in the left of Fig.\ref{fig:duration_result}(a), during the intervention stage, \projectname resulted in 36.0 $\pm$ 22.6\% of the duration compared to the pre-intervention stage (\ie a reduction of 64.0\% of the target undesirable action), and the baseline system led to 65.0 $\pm$ 47.5\% of the duration (\ie a reduction of 35.0\%).
We fitted a GLMM to compare the two intervention methods.
Our results revealed the significant difference between the two methods: \projectname significantly outperformed the baseline by 29.0\% more reduction of the target undesirable action ($\chi^2_1=6.32, p < .05$). Meanwhile, the number of notifications does not show significance ($\chi^2_1=0.53, p = 0.47$). These results suggest that the advantage of \projectname was mainly attributed to the AI-based intervention method.

In addition, although our post-intervention stage was short, both methods showed promising signals of a potential lasting effect when the intervention was gone (13.9 $\pm$ 16.8\% for the \projectname; 37.7 $\pm$ 37.2\% for the baseline), as shown in the right of Fig.\ref{fig:duration_result}(a).
We fitted another GLMM on the post-intervention data. The results also indicate the significance of the intervention method ($\chi^2_1=10.04, p<0.01$), but not the number of notifications ($\chi^2_1=0.12, p=0.73$).
This is consistent with the result of the intervention stage, further demonstrating the superior performance of \projectname over the baseline method.



\subsubsection{Intervention Effectiveness over Time.}
To investigate changes in the duration of target action during the study session, we visualize the change of participants' target action duration throughout the study (see \autoref{fig:duration_result}(b)).
Both intervention methods showed a clear and significant decreasing trend once participants entered the intervention stage.
The fitted lines in \autoref{fig:duration_result}(b) indicate that \projectname achieved more duration reduction ($m=-4.8\%$ per 10-minute) compared to the baseline ($m=-4.1\%$) over the intervention session.
In particular, \projectname had a more rapid initial decrease and maintained consistently lower levels throughout the rest of the session compared to the rule-based baseline.
Overall, \projectname demonstrated stronger cumulative effects.

\subsubsection{Difference across Task Types.}
\begin{figure*}[]
\centering 
\includegraphics[width=0.7\linewidth]{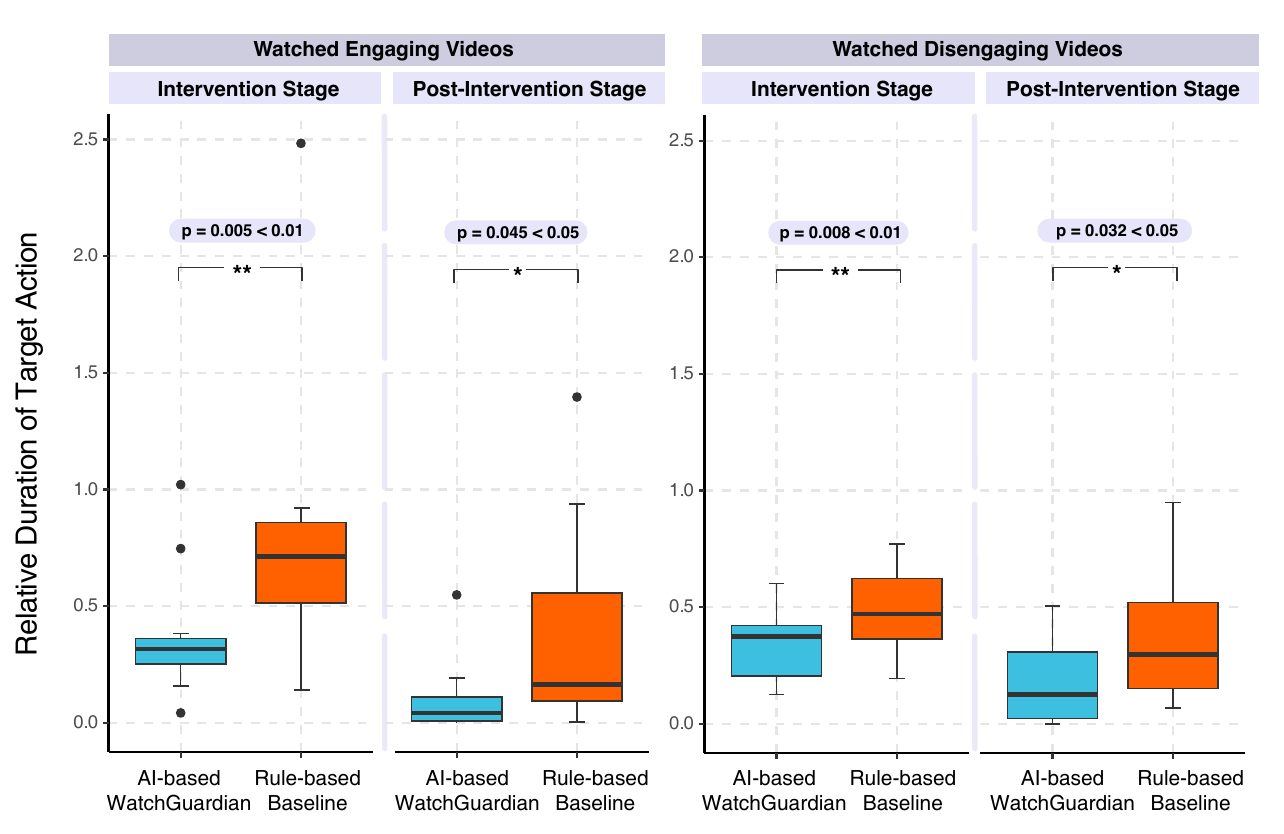}
\caption{
(a) Relative duration of target action for participants who watched engaging videos.
(b) Relative duration of target action for participants who watched disengaging videos.
}
\label{fig:duration_result_breakdown}
\Description{}
\end{figure*}

During the study, we asked participants to pick their own preferred task types between watching engaging (N=11) or disengaging videos (N=10).
\autoref{fig:duration_result_breakdown} presents the breakdown of the task type in \autoref{fig:duration_result}(a).
We fitted GLMMs with task type as another main factor and observed a marginal significance of the interaction between the intervention method and the task type ($\chi^2_1=3.27, p=0.07<0.1$). This was only during the intervention stage, but not the post-intervention stage.
\autoref{fig:duration_result_breakdown}(a) and (b) indicate that the advantage of \projectname during the intervention stage was more salient when participants were watching engaging videos ($\Delta=42.3\pm49.6\%$) compared to when they were watching disengaging videos ($\Delta=14.2\pm22.5\%$).
This could be due to the fact that participants were more interruptable or receptive in less engaging tasks~\cite{pielot2017beyond,mishra_detecting_2021,choi_multi-stage_2019}, thus even a basic rule-based intervention could effectively reduce the target actions. However, in more engaging tasks, accurate and just-in-time reminders are more effective than basic ones.

\subsubsection{Survey Outcomes.}
In addition to the objective measurement, we also compare participants' subjective reports on the SUS, WAI, and the change of the habit strength.
Overall, participants reported that \projectname had better usability (SUS: 73.3$\pm$12.8) than the baseline (66.8$\pm$15.9), with significance through a Wilcoxon rank sum test ($p<0.05$).
\projectname achieved a SUS score over 70, indicating acceptable usability. In both methods, false positive notifications were inevitable and could introduce participants' confusion or surprise, which could explain the subpar SUS scores in general.

Interestingly, the results of WAI and habit strength did not indicate such a difference. Participants had similar reports of the relationship with the system (WAI score: 42.1$\pm$7.1 for the \projectname \vs 41.9$\pm$9.3 for the baseline, $p=0.58$).
The change of habit strength between the pre- and post-intervention stages is also minimal ($\Delta$ of habit strength score: $-4.2\pm5.5$ for the \projectname \vs $-5.5\pm6.6$ for the baseline, $p=0.25$).
This was probably due to the fact that the intervention sessions were not long enough to form a long-term alliance between users and the system, or to influence longitudinal behaviors or habits.
Our qualitative results from semi-structured interviews provide more nuanced insights into these results.

\subsection{Qualitative Results}
\label{sub:intervention_evaluation:qualitative_results}
All interviews were recorded and transcribed. We adopted a simple content analysis framework~\cite{prior2014content}. One author took extensive notes during the interviews, went through the scripts to categorize themes and count their frequency, and discussed with two other authors until convergence.
We summarize our key findings below.


\subsubsection{Perception of AI-powered Intervention}
Multiple participants reported that the AI-powered system possessed a sense of presence or "\textit{having a soul}". For instance, P10 noted, "\textit{[\projectname{}] resembles a habit instructor, or even like my mom... who would gently remind me when I scratch my head.}"
P18 remarked, "\textit{This system seems to read my mind, anticipating when I'm about to bite my lips and reminding me just in time. Sometimes I felt like I was sneaking around when making these actions.}"
Compared to the rule-based condition, \projectname's interventions appeared to foster greater self-reflection among users.
Notably, P19 even perceived the AI's reminders as rewards: "\textit{After being caught [touching my face] several times initially, I managed to control myself for a while. Then, even if the system reminded me again, I felt it was affirming my progress, like receiving a reward.}"
In contrast, the rule-based condition yielded opposite effects, "\textit{This mode of notification felt random to me - it was just like a machine}" (P02).

However, some participants also had a negative experience with \projectname, especially when it did not detect the actions accurately (mostly false positive). For example, P08 mentioned that \projectname had limited impact, and that they also felt a sense of distrust. "\textit{At first, when it reported errors a few times, I tried to look for reasons elsewhere. But it kept making mistakes, which became frustrating. When it occasionally got something right, I thought it was just luck!}" Participants could lose trust in \projectname when the system made mistakes at the beginning of their interaction.
This is supported by prior research in other human-AI interaction systems~\cite{swaroop2024accuracy,jacobs2021designing}.

\subsubsection{Illusory Amplification of Intervention Strength}
We noticed a surprisingly interesting phenomenon: 
Several users (P09, P10, and P16) reported that the vibration strength of the AI-based intervention in \projectname felt stronger than that of the rule-based intervention. However, the vibration setup was identical in the two sessions.
Even after we explained the specific intervention methods after the two study sessions, P16 stated, "\textit{Not only did I subjectively feel that Mode B [our \projectname method] gave me a stronger sense of motion restraint, but it also seemed to vibrate more intensely. Are you sure it's really the same setting?}"
This indicated that participants might develop an illusory or distorted perception of the intervention's strength when the interventions were delivered just-in-time.
We discuss this more in Sec.~\ref{sub:discussion:distorted}.

\subsubsection{Diverse Patterns of Human-AI Collaborative Relationship}
Users exhibited diverse patterns of engagement with the AI system. Some participants demonstrated adaptive behavioral modification in response to \projectname's reminders. As P14 described, "\textit{Every time I shook my leg, it would remind me, which made me increasingly hesitant to move}". This was aligned with our original design goal of introducing AI-powered JITI.

Other than reducing the target actions, we also observed other behavior patterns. One pattern emerged where participants developed an interesting competitive relationship with the AI for user agency. For instance, P8 articulated this sentiment: "\textit{I wanted to compete with it - I tried to resist the urge just so it wouldn't catch me.}" This competitive spirit evolved into experimental behavior for some users, who attempted to understand and control the system's underlying logic. P18's experience exemplified this progression: "\textit{Initially, I felt caught red-handed with every reminder. Later, I noticed it wouldn't always detect my subtle movements, so I started experimenting with the notification logic, trying to gain control over the reminders. Eventually, though, I made peace with it and lost the urge to perform the action altogether.}" These participants wanted to gain better agency in this human-AI relationship.

In addition, some participants developed playful interactions with the system, treating it as an engaging companion rather than a mere monitoring tool. For example, P17 shared: "\textit{When the video was boring... I just wanted to goof around a bit. This thing was actually keeping an eye on me, so I'd mess with it for fun, play around with it, and boom - it would react right away. Kinda helped wake me up a bit? It was basically like playing a game.}"
Overall, these diverse patterns between users and \projectname suggest a set of potential collaborative relationships between the two sides. We discuss this finding in Sec.~\ref{sub:discussion:collaboration}.

\section{Discussion}
\label{sec:discussion}
In this work, we propose to leverage few-shot learning to enable users to self-define personal undesirable actions for personalized intervention on smartwatches.
We developed a three-stage pipeline that began with a self-supervised, pre-trained IMU model for robust feature extraction, then fine-tuned it for accurate human activity recognition, and finally enhanced it with data augmentation and synthesis that enabled rapid customization of new user-defined actions using only a small number of examples. 
We implemented this pipeline on a smartwatch as a real-time intervention system, \projectname, and demonstrated its effectiveness and advantages over the rule-based method through a multi-hour user study.
In this section, we discuss some interesting takeaways from our study, together with our vision of how \projectname can be generally applied to other health domains. We also briefly summarize the limitations of our work.

\subsection{Distorted Perception with AI-powered Intervention}
\label{sub:discussion:distorted}
During the study, we observed an interesting phenomenon where some participants developed a distorted perception towards their own actions or the intervention (see Sec.~\ref{sub:intervention_evaluation:qualitative_results}).
For instance, several participants felt \projectname's vibrations were stronger than the baseline (yet the actual strength of vibration remained constant), and some felt they did the target actions more frequently with \projectname (yet the objective data indicated otherwise).
There are several potential interpretations of such interesting observations.
The distorted perception might be caused by participants' heightened awareness of the AI-guided interventions: because \projectname more accurately and promptly caught the target actions, users started to pay extra and prolonged attention to any intervention. This could leave a stronger impression on them, and subsequently, they found it stronger or more frequent.
Another potential explanation is that the participants, often associating their personal and idiosyncratic undesirable actions with ``wrong-doing'' and thus responding with negative emotions, might have subconsciously perceived their undesirable actions as being more frequent due to the \projectname's more precise and timely feedback eliciting stronger negative emotions. This, combined with an emotional interpretation of being 'corrected', may have amplified their perception of the intervention's intensity (vibration strength) and created the mistaken impression of performing these actions excessively.

Meanwhile, it is an interesting open question of how long such perception will last from a longitudinal intervention perspective. Depending on the cases, the growing self-awareness and/or reliability of AI could potentially assist users in building a long-term habit to reduce the target action, or on the contrary, the effect may fade away due to the AI intervention method no longer being novel or enticing.
Future work can explore the lasting effect of the intervention, the corresponding perception, as well as user engagement in a long-term, field-based intervention study.~\cite{middleton2013long, short2018measuring, wei2020design}.

\subsection{Towards Human-AI Collaborative Interventions}
\label{sub:discussion:collaboration}
Users' mental models of \projectname varied significantly. Some viewed it as a passive watchdog, and some viewed it as a playful interactive system, while others sought to take greater agency in the moment of intervention delivery.
Our findings show the potential for and benefit of developing a collaborative relationship between humans and AI for behavioral intervention.
An AI system can provide appropriate support to users and help them achieve effective intervention outcomes.
Such collaboration is closely relevant to the vision of just-in-time adaptive interventions (JITAIs)~\cite{nahum-shani_translating_2021, nahum2018just}, where the delivery timing and methods of intervention are designed to be dynamically adapting to an individual's internal state and surrounding context.

For instance, for users who see the system as a passive monitor, the system can provide the option for them to configure the frequency and style of intervention (\eg higher/lower-intensity vibrations or consolidated notifications), ensuring the AI remains in the background but still provides supportive nudges.
Taking one step further, the AI system may analyze user behavior over time and suggest new setups or goals for users with transparency (\eg transitioning from nail-biting to managing stress). Users can accept, modify, or reject these suggestions, creating a dialogue where AI acts as a coach or collaborator rather than a rigid enforcer of predefined behaviors.
Meanwhile, for those who see AI as a proactive system, one promising avenue is to incorporate user feedback into the AI's learning process~\cite{orzikulova2024time2stop}. Users can label the AI's predictions as accurate or not, which could serve as input for the model to further adapt to the user and improve performance over time (\eg through reinforcement learning).
Combined with contextual information that can potentially be inferred from sensors~\cite{xu2023globem}, such feedback can enable more precise, context-sensitive and personalized JITIs.
In addition, the system would periodically prompt users to reassess their goals and update intervention targets, ensuring long-term relevance and efficacy.

It is noteworthy that such a human-AI collaboration paradigm needs to follow the principles of transparency and ethical design.
Other than the options mentioned above, namely custom configurations and continuous feedback, users should have visibility into the system's functionality and action logic regardless of the specific collaboration setup. This is important to provide users with agency and build their trust in the system.

\subsection{Beyond Smartwatch and Broader Customization}
In this work, our real-time intervention was implemented on a smartwatch. However, our proposed idea of empowering users to define any personal action and achieve a personalized intervention system can be more broadly applied to other domains.
Instead of relying solely on a watch-based IMU, we can explore other body-based sensor arrays (\eg headbands, rings, or joint sensors) to capture a more diverse range of behaviors in real time.
This would enable the system to accommodate a wide variety of undesirable actions or habits, such as posture corrections and fidgeting management.
In addition, beyond physical interventions, future customization can also delve into psychological or mental health support.
For instance, individuals dealing with obsessive-compulsive disorder (OCD) or other habitual thought/action patterns could define personal triggers (\eg a particular repetitive motion or behavioral cue) and receive timely AI-driven interventions.
Such holistic approaches highlight the flexibility and scalability of our pipeline.
By enabling user-defined actions, we open up possibilities for long-term and effective management of both physical and psychological well-being using a multitude of wearable and sensor-based platforms.

\subsection{Limitations}

Despite \projectname's positive outcome and the promising insights generated, we recognize some limitations in our study design.
As mentioned above, our current model relies solely on accelerometer data for action recognition, which may limit its ability to capture the full range of motion characteristics or other physiology. Future work can explore additional sensing modalities, such as gyroscope, photoplethysmography (PPG), joint locations, to enhance the accuracy and robustness of action recognition. 
Besides, the study was conducted with a relatively small number of participants and a limited set of actions, which may not fully capture the variability and diversity of human activities in real-world scenarios \cite{trapp2015individual, narayanan2013behavioral}.
Additionally, although we tried to simulate real-life scenarios, our intervention study was conducted over a limited duration and in controlled experimental settings, which may not fully reflect the complexities and dynamics of real-life environments. 
Real-world contexts introduce factors such as environmental noise, varying sensor placements, and user behavior changes over time \cite{trapp2015individual,truong2015deployment,mejia2023enhancing,mills2022development}, which were not thoroughly simulated in this study. Future research should conduct longitudinal field experiments with real-world deployment of the system.

\section{Conclusion}
\label{sec:conclusion}
In this study, we introduced \projectname, a smartwatch-based intervention system that empowers users to define and reduce their own personal and idiosyncratic undesirable actions.
\projectname employs a three-stage, few-shot learning pipeline to recognize newly defined actions with only a small number of samples.
Through an extensive evaluation of the model's offline performance and an intervention study, our findings demonstrate that \projectname achieves robust and data-efficient action recognition but also significantly decreases users’ undesirable actions compared to a rule-based baseline.
Our findings underscore the potential of personalized, AI-driven JITIs for individuals seeking to mitigate personal habits and behaviors.
We envision that our work offers a versatile foundation for creating broader, user-defined intervention systems that leverage advanced AI solutions to accommodate an ever-growing spectrum of personal needs.


\bibliographystyle{ACM-Ref-Format}
\bibliography{
bib/Orson/BehaviorIntervention,
bib/Orson/HumanComputerInteraction,
bib/Orson/Modeling_Behavior-General,
bib/Orson/MachineLearning,
bib/Orson/OrsonPublication,
bib/ly/BehaviorIntervention,
bib/ly/MachineLearning,
bib/ly/Dataset,
bib/ly/others,
bib/others
}

\appendix
\newpage
\label{appendix:}
\section{Few-shot Pipeline Performance with Different Numbers of Actions and Shots}

\renewcommand{\arraystretch}{1.28}

\begin{table}[htbp]
\centering
\resizebox{1\textwidth}{!}{
\begin{tabular}[]{c|c|cccc|cccc}
\toprule
\multirow{2}{*}{\textbf{Actions}} & \multirow{2}{*}{\textbf{Shots}}  & \multicolumn{4}{c}{\textbf{Window-level}}      & \multicolumn{4}{c}{\textbf{Action-level}}       \\ \cline{3-10}
 & & \textbf{Acc}   & \textbf{Prec}  & \textbf{Rec}   & \textbf{F1}    & \textbf{Acc}   & \textbf{Prec}  & \textbf{Rec}   & \textbf{F1}      \\ \hline
\multirow{10}{*}{1} 
  & 1  & 0.614{\tiny$\pm$0.006} & 0.700{\tiny$\pm$0.007} & 0.614{\tiny$\pm$0.006} & 0.571{\tiny$\pm$0.008} & 0.768{\tiny$\pm$0.007} & 0.810{\tiny$\pm$0.007} & 0.768{\tiny$\pm$0.007} & 0.748{\tiny$\pm$0.009} \\
  & 2  & 0.628{\tiny$\pm$0.005} & 0.710{\tiny$\pm$0.006} & 0.628{\tiny$\pm$0.005} & 0.620{\tiny$\pm$0.006} & 0.800{\tiny$\pm$0.006} & 0.835{\tiny$\pm$0.006} & 0.800{\tiny$\pm$0.006} & 0.810{\tiny$\pm$0.006} \\
  & 3  & 0.658{\tiny$\pm$0.005} & 0.736{\tiny$\pm$0.005} & 0.658{\tiny$\pm$0.005} & 0.634{\tiny$\pm$0.006} & 0.832{\tiny$\pm$0.006} & 0.860{\tiny$\pm$0.005} & 0.832{\tiny$\pm$0.006} & 0.825{\tiny$\pm$0.006} \\
  & 4  & 0.664{\tiny$\pm$0.005} & 0.743{\tiny$\pm$0.005} & 0.664{\tiny$\pm$0.005} & 0.641{\tiny$\pm$0.006} & 0.839{\tiny$\pm$0.005} & 0.866{\tiny$\pm$0.005} & 0.839{\tiny$\pm$0.005} & 0.833{\tiny$\pm$0.006} \\
  & 5  & 0.670{\tiny$\pm$0.005} & 0.746{\tiny$\pm$0.005} & 0.670{\tiny$\pm$0.005} & 0.648{\tiny$\pm$0.006} & 0.847{\tiny$\pm$0.005} & 0.871{\tiny$\pm$0.005} & 0.847{\tiny$\pm$0.005} & 0.842{\tiny$\pm$0.006} \\
  & 6  & 0.676{\tiny$\pm$0.005} & 0.749{\tiny$\pm$0.005} & 0.676{\tiny$\pm$0.005} & 0.657{\tiny$\pm$0.006} & 0.856{\tiny$\pm$0.005} & 0.879{\tiny$\pm$0.005} & 0.856{\tiny$\pm$0.005} & 0.852{\tiny$\pm$0.006} \\
  & 7  & 0.685{\tiny$\pm$0.005} & 0.755{\tiny$\pm$0.005} & 0.685{\tiny$\pm$0.005} & 0.667{\tiny$\pm$0.006} & 0.864{\tiny$\pm$0.005} & 0.883{\tiny$\pm$0.005} & 0.864{\tiny$\pm$0.005} & 0.860{\tiny$\pm$0.006} \\
  & 8  & 0.691{\tiny$\pm$0.005} & 0.758{\tiny$\pm$0.005} & 0.691{\tiny$\pm$0.005} & 0.674{\tiny$\pm$0.006} & 0.869{\tiny$\pm$0.005} & 0.887{\tiny$\pm$0.005} & 0.869{\tiny$\pm$0.005} & 0.865{\tiny$\pm$0.006} \\
  & 9  & 0.696{\tiny$\pm$0.005} & 0.760{\tiny$\pm$0.005} & 0.696{\tiny$\pm$0.005} & 0.680{\tiny$\pm$0.006} & 0.870{\tiny$\pm$0.005} & 0.885{\tiny$\pm$0.005} & 0.870{\tiny$\pm$0.005} & 0.865{\tiny$\pm$0.006} \\
  & 10 & 0.702{\tiny$\pm$0.005} & 0.763{\tiny$\pm$0.005} & 0.702{\tiny$\pm$0.005} & 0.688{\tiny$\pm$0.006} & 0.877{\tiny$\pm$0.005} & 0.890{\tiny$\pm$0.005} & 0.877{\tiny$\pm$0.005} & 0.873{\tiny$\pm$0.006} \\
\midrule
\multirow{10}{*}{2} 
  & 1  & 0.578{\tiny$\pm$0.003} & 0.622{\tiny$\pm$0.004} & 0.578{\tiny$\pm$0.003} & 0.521{\tiny$\pm$0.004} & 0.701{\tiny$\pm$0.004} & 0.717{\tiny$\pm$0.004} & 0.701{\tiny$\pm$0.004} & 0.667{\tiny$\pm$0.004} \\
  & 2  & 0.589{\tiny$\pm$0.003} & 0.635{\tiny$\pm$0.004} & 0.589{\tiny$\pm$0.003} & 0.545{\tiny$\pm$0.004} & 0.745{\tiny$\pm$0.004} & 0.760{\tiny$\pm$0.004} & 0.745{\tiny$\pm$0.004} & 0.730{\tiny$\pm$0.004} \\
  & 3  & 0.639{\tiny$\pm$0.003} & 0.684{\tiny$\pm$0.003} & 0.639{\tiny$\pm$0.003} & 0.608{\tiny$\pm$0.003} & 0.783{\tiny$\pm$0.003} & 0.797{\tiny$\pm$0.003} & 0.783{\tiny$\pm$0.003} & 0.770{\tiny$\pm$0.003} \\
  & 4  & 0.653{\tiny$\pm$0.003} & 0.696{\tiny$\pm$0.003} & 0.653{\tiny$\pm$0.003} & 0.625{\tiny$\pm$0.003} & 0.803{\tiny$\pm$0.003} & 0.815{\tiny$\pm$0.003} & 0.803{\tiny$\pm$0.003} & 0.792{\tiny$\pm$0.003} \\
  & 5  & 0.667{\tiny$\pm$0.003} & 0.707{\tiny$\pm$0.003} & 0.667{\tiny$\pm$0.003} & 0.643{\tiny$\pm$0.003} & 0.817{\tiny$\pm$0.003} & 0.827{\tiny$\pm$0.003} & 0.817{\tiny$\pm$0.003} & 0.808{\tiny$\pm$0.003} \\
  & 6  & 0.679{\tiny$\pm$0.003} & 0.717{\tiny$\pm$0.003} & 0.679{\tiny$\pm$0.003} & 0.659{\tiny$\pm$0.003} & 0.829{\tiny$\pm$0.003} & 0.838{\tiny$\pm$0.003} & 0.829{\tiny$\pm$0.003} & 0.821{\tiny$\pm$0.003} \\
  & 7  & 0.690{\tiny$\pm$0.003} & 0.725{\tiny$\pm$0.003} & 0.690{\tiny$\pm$0.003} & 0.672{\tiny$\pm$0.003} & 0.838{\tiny$\pm$0.003} & 0.847{\tiny$\pm$0.003} & 0.838{\tiny$\pm$0.003} & 0.832{\tiny$\pm$0.003} \\
  & 8  & 0.699{\tiny$\pm$0.003} & 0.731{\tiny$\pm$0.003} & 0.699{\tiny$\pm$0.003} & 0.682{\tiny$\pm$0.003} & 0.847{\tiny$\pm$0.003} & 0.854{\tiny$\pm$0.003} & 0.847{\tiny$\pm$0.003} & 0.841{\tiny$\pm$0.003} \\
  & 9  & 0.704{\tiny$\pm$0.003} & 0.735{\tiny$\pm$0.003} & 0.704{\tiny$\pm$0.003} & 0.689{\tiny$\pm$0.003} & 0.850{\tiny$\pm$0.003} & 0.858{\tiny$\pm$0.003} & 0.850{\tiny$\pm$0.003} & 0.845{\tiny$\pm$0.003} \\
  & 10 & 0.708{\tiny$\pm$0.003} & 0.738{\tiny$\pm$0.003} & 0.708{\tiny$\pm$0.003} & 0.695{\tiny$\pm$0.003} & 0.854{\tiny$\pm$0.003} & 0.862{\tiny$\pm$0.003} & 0.854{\tiny$\pm$0.003} & 0.849{\tiny$\pm$0.003} \\
\midrule
\multirow{10}{*}{3} 
  & 1  & 0.566{\tiny$\pm$0.002} & 0.572{\tiny$\pm$0.003} & 0.566{\tiny$\pm$0.002} & 0.511{\tiny$\pm$0.003} & 0.670{\tiny$\pm$0.003} & 0.667{\tiny$\pm$0.004} & 0.670{\tiny$\pm$0.003} & 0.636{\tiny$\pm$0.003} \\
  & 2  & 0.587{\tiny$\pm$0.002} & 0.605{\tiny$\pm$0.003} & 0.587{\tiny$\pm$0.002} & 0.555{\tiny$\pm$0.003} & 0.723{\tiny$\pm$0.003} & 0.730{\tiny$\pm$0.003} & 0.723{\tiny$\pm$0.003} & 0.710{\tiny$\pm$0.003} \\
  & 3  & 0.630{\tiny$\pm$0.002} & 0.654{\tiny$\pm$0.003} & 0.630{\tiny$\pm$0.002} & 0.596{\tiny$\pm$0.003} & 0.758{\tiny$\pm$0.002} & 0.766{\tiny$\pm$0.003} & 0.758{\tiny$\pm$0.002} & 0.742{\tiny$\pm$0.003} \\
  & 4  & 0.647{\tiny$\pm$0.002} & 0.671{\tiny$\pm$0.003} & 0.647{\tiny$\pm$0.002} & 0.617{\tiny$\pm$0.003} & 0.777{\tiny$\pm$0.002} & 0.785{\tiny$\pm$0.003} & 0.777{\tiny$\pm$0.002} & 0.764{\tiny$\pm$0.003} \\
  & 5  & 0.660{\tiny$\pm$0.002} & 0.683{\tiny$\pm$0.002} & 0.660{\tiny$\pm$0.002} & 0.635{\tiny$\pm$0.003} & 0.791{\tiny$\pm$0.002} & 0.799{\tiny$\pm$0.003} & 0.791{\tiny$\pm$0.002} & 0.781{\tiny$\pm$0.003} \\
  & 6  & 0.673{\tiny$\pm$0.002} & 0.694{\tiny$\pm$0.002} & 0.673{\tiny$\pm$0.002} & 0.650{\tiny$\pm$0.003} & 0.804{\tiny$\pm$0.002} & 0.812{\tiny$\pm$0.002} & 0.804{\tiny$\pm$0.002} & 0.795{\tiny$\pm$0.003} \\
  & 7  & 0.682{\tiny$\pm$0.002} & 0.701{\tiny$\pm$0.002} & 0.682{\tiny$\pm$0.002} & 0.662{\tiny$\pm$0.003} & 0.811{\tiny$\pm$0.002} & 0.819{\tiny$\pm$0.002} & 0.811{\tiny$\pm$0.002} & 0.803{\tiny$\pm$0.002} \\
  & 8  & 0.690{\tiny$\pm$0.002} & 0.709{\tiny$\pm$0.002} & 0.690{\tiny$\pm$0.002} & 0.672{\tiny$\pm$0.003} & 0.817{\tiny$\pm$0.002} & 0.826{\tiny$\pm$0.002} & 0.817{\tiny$\pm$0.002} & 0.811{\tiny$\pm$0.002} \\
  & 9  & 0.697{\tiny$\pm$0.002} & 0.714{\tiny$\pm$0.002} & 0.697{\tiny$\pm$0.002} & 0.680{\tiny$\pm$0.003} & 0.823{\tiny$\pm$0.002} & 0.831{\tiny$\pm$0.002} & 0.823{\tiny$\pm$0.002} & 0.818{\tiny$\pm$0.002} \\
  & 10 & 0.701{\tiny$\pm$0.002} & 0.718{\tiny$\pm$0.002} & 0.701{\tiny$\pm$0.002} & 0.686{\tiny$\pm$0.003} & 0.826{\tiny$\pm$0.002} & 0.835{\tiny$\pm$0.002} & 0.826{\tiny$\pm$0.002} & 0.822{\tiny$\pm$0.002} \\
\bottomrule
\end{tabular}
}
\caption{Prediction Performance with Different Number of Actions and Shots. (Action Number 1--3)}
\end{table}

\newpage

\begin{table}[htbp]
\centering
\resizebox{1\textwidth}{!}{
\begin{tabular}[]{c|c|cccc|cccc}
\toprule
\multirow{2}{*}{\textbf{Actions}} & \multirow{2}{*}{\textbf{Shots}}  & \multicolumn{4}{c}{\textbf{Window-level}}      & \multicolumn{4}{c}{\textbf{Action-level}}       \\ \cline{3-10}
 & & \textbf{Acc}   & \textbf{Prec}  & \textbf{Rec}   & \textbf{F1}    & \textbf{Acc}   & \textbf{Prec}  & \textbf{Rec}   & \textbf{F1}      \\ \hline
\multirow{10}{*}{4} 
  & 1  & 0.557{\tiny$\pm$0.002} & 0.553{\tiny$\pm$0.003} & 0.557{\tiny$\pm$0.002} & 0.501{\tiny$\pm$0.003} & 0.659{\tiny$\pm$0.003} & 0.656{\tiny$\pm$0.003} & 0.659{\tiny$\pm$0.003} & 0.627{\tiny$\pm$0.003} \\
  & 2  & 0.589{\tiny$\pm$0.002} & 0.608{\tiny$\pm$0.003} & 0.589{\tiny$\pm$0.002} & 0.555{\tiny$\pm$0.003} & 0.723{\tiny$\pm$0.003} & 0.730{\tiny$\pm$0.003} & 0.723{\tiny$\pm$0.003} & 0.710{\tiny$\pm$0.003} \\
  & 3  & 0.619{\tiny$\pm$0.002} & 0.635{\tiny$\pm$0.003} & 0.619{\tiny$\pm$0.002} & 0.583{\tiny$\pm$0.003} & 0.739{\tiny$\pm$0.003} & 0.745{\tiny$\pm$0.003} & 0.739{\tiny$\pm$0.003} & 0.723{\tiny$\pm$0.003} \\
  & 4  & 0.635{\tiny$\pm$0.002} & 0.651{\tiny$\pm$0.003} & 0.635{\tiny$\pm$0.002} & 0.604{\tiny$\pm$0.003} & 0.757{\tiny$\pm$0.003} & 0.766{\tiny$\pm$0.003} & 0.757{\tiny$\pm$0.003} & 0.745{\tiny$\pm$0.003} \\
  & 5  & 0.647{\tiny$\pm$0.002} & 0.663{\tiny$\pm$0.003} & 0.647{\tiny$\pm$0.002} & 0.621{\tiny$\pm$0.003} & 0.770{\tiny$\pm$0.002} & 0.778{\tiny$\pm$0.003} & 0.770{\tiny$\pm$0.002} & 0.761{\tiny$\pm$0.003} \\
  & 6  & 0.658{\tiny$\pm$0.002} & 0.673{\tiny$\pm$0.003} & 0.658{\tiny$\pm$0.002} & 0.635{\tiny$\pm$0.003} & 0.782{\tiny$\pm$0.002} & 0.791{\tiny$\pm$0.002} & 0.782{\tiny$\pm$0.002} & 0.775{\tiny$\pm$0.003} \\
  & 7  & 0.667{\tiny$\pm$0.002} & 0.681{\tiny$\pm$0.003} & 0.667{\tiny$\pm$0.002} & 0.646{\tiny$\pm$0.003} & 0.790{\tiny$\pm$0.002} & 0.799{\tiny$\pm$0.002} & 0.790{\tiny$\pm$0.002} & 0.783{\tiny$\pm$0.002} \\
  & 8  & 0.676{\tiny$\pm$0.002} & 0.689{\tiny$\pm$0.003} & 0.676{\tiny$\pm$0.002} & 0.657{\tiny$\pm$0.003} & 0.796{\tiny$\pm$0.002} & 0.805{\tiny$\pm$0.002} & 0.796{\tiny$\pm$0.002} & 0.791{\tiny$\pm$0.002} \\
  & 9  & 0.684{\tiny$\pm$0.002} & 0.697{\tiny$\pm$0.003} & 0.684{\tiny$\pm$0.002} & 0.667{\tiny$\pm$0.003} & 0.803{\tiny$\pm$0.002} & 0.814{\tiny$\pm$0.002} & 0.803{\tiny$\pm$0.002} & 0.800{\tiny$\pm$0.002} \\
  & 10 & 0.689{\tiny$\pm$0.002} & 0.701{\tiny$\pm$0.003} & 0.689{\tiny$\pm$0.002} & 0.673{\tiny$\pm$0.003} & 0.807{\tiny$\pm$0.002} & 0.817{\tiny$\pm$0.002} & 0.807{\tiny$\pm$0.002} & 0.803{\tiny$\pm$0.002} \\
\midrule
\multirow{10}{*}{5} 
  & 1  & 0.542{\tiny$\pm$0.003} & 0.524{\tiny$\pm$0.004} & 0.542{\tiny$\pm$0.003} & 0.481{\tiny$\pm$0.004} & 0.636{\tiny$\pm$0.004} & 0.618{\tiny$\pm$0.005} & 0.636{\tiny$\pm$0.004} & 0.600{\tiny$\pm$0.005} \\
  & 2  & 0.584{\tiny$\pm$0.003} & 0.581{\tiny$\pm$0.004} & 0.584{\tiny$\pm$0.003} & 0.538{\tiny$\pm$0.004} & 0.691{\tiny$\pm$0.004} & 0.689{\tiny$\pm$0.004} & 0.691{\tiny$\pm$0.004} & 0.668{\tiny$\pm$0.004} \\
  & 3  & 0.610{\tiny$\pm$0.003} & 0.612{\tiny$\pm$0.004} & 0.610{\tiny$\pm$0.003} & 0.573{\tiny$\pm$0.004} & 0.722{\tiny$\pm$0.004} & 0.724{\tiny$\pm$0.004} & 0.722{\tiny$\pm$0.004} & 0.705{\tiny$\pm$0.004} \\
  & 4  & 0.629{\tiny$\pm$0.004} & 0.635{\tiny$\pm$0.004} & 0.629{\tiny$\pm$0.004} & 0.597{\tiny$\pm$0.004} & 0.741{\tiny$\pm$0.004} & 0.746{\tiny$\pm$0.004} & 0.741{\tiny$\pm$0.004} & 0.728{\tiny$\pm$0.004} \\
  & 5  & 0.643{\tiny$\pm$0.004} & 0.649{\tiny$\pm$0.004} & 0.643{\tiny$\pm$0.004} & 0.614{\tiny$\pm$0.004} & 0.755{\tiny$\pm$0.004} & 0.762{\tiny$\pm$0.004} & 0.755{\tiny$\pm$0.004} & 0.745{\tiny$\pm$0.004} \\
  & 6  & 0.654{\tiny$\pm$0.004} & 0.661{\tiny$\pm$0.004} & 0.654{\tiny$\pm$0.004} & 0.629{\tiny$\pm$0.004} & 0.767{\tiny$\pm$0.004} & 0.774{\tiny$\pm$0.004} & 0.767{\tiny$\pm$0.004} & 0.759{\tiny$\pm$0.004} \\
  & 7  & 0.663{\tiny$\pm$0.004} & 0.670{\tiny$\pm$0.004} & 0.663{\tiny$\pm$0.004} & 0.640{\tiny$\pm$0.004} & 0.774{\tiny$\pm$0.004} & 0.782{\tiny$\pm$0.004} & 0.774{\tiny$\pm$0.004} & 0.768{\tiny$\pm$0.004} \\
  & 8  & 0.672{\tiny$\pm$0.004} & 0.679{\tiny$\pm$0.004} & 0.672{\tiny$\pm$0.004} & 0.651{\tiny$\pm$0.004} & 0.785{\tiny$\pm$0.004} & 0.793{\tiny$\pm$0.004} & 0.785{\tiny$\pm$0.004} & 0.779{\tiny$\pm$0.004} \\
  & 9  & 0.678{\tiny$\pm$0.004} & 0.685{\tiny$\pm$0.004} & 0.678{\tiny$\pm$0.004} & 0.659{\tiny$\pm$0.004} & 0.789{\tiny$\pm$0.004} & 0.798{\tiny$\pm$0.004} & 0.789{\tiny$\pm$0.004} & 0.784{\tiny$\pm$0.004} \\
  & 10 & 0.684{\tiny$\pm$0.004} & 0.690{\tiny$\pm$0.004} & 0.684{\tiny$\pm$0.004} & 0.666{\tiny$\pm$0.004} & 0.795{\tiny$\pm$0.004} & 0.803{\tiny$\pm$0.004} & 0.795{\tiny$\pm$0.004} & 0.790{\tiny$\pm$0.004} \\
\midrule
\multirow{10}{*}{6} 
  & 1  & 0.533{\tiny$\pm$0.009} & 0.510{\tiny$\pm$0.010} & 0.533{\tiny$\pm$0.009} & 0.479{\tiny$\pm$0.010} & 0.617{\tiny$\pm$0.010} & 0.613{\tiny$\pm$0.011} & 0.617{\tiny$\pm$0.010} & 0.589{\tiny$\pm$0.011} \\
  & 2  & 0.579{\tiny$\pm$0.008} & 0.565{\tiny$\pm$0.010} & 0.579{\tiny$\pm$0.008} & 0.533{\tiny$\pm$0.010} & 0.672{\tiny$\pm$0.009} & 0.668{\tiny$\pm$0.011} & 0.672{\tiny$\pm$0.009} & 0.650{\tiny$\pm$0.010} \\
  & 3  & 0.599{\tiny$\pm$0.009} & 0.592{\tiny$\pm$0.010} & 0.599{\tiny$\pm$0.009} & 0.561{\tiny$\pm$0.010} & 0.703{\tiny$\pm$0.010} & 0.706{\tiny$\pm$0.011} & 0.703{\tiny$\pm$0.010} & 0.686{\tiny$\pm$0.011} \\
  & 4  & 0.617{\tiny$\pm$0.009} & 0.612{\tiny$\pm$0.011} & 0.617{\tiny$\pm$0.009} & 0.583{\tiny$\pm$0.011} & 0.720{\tiny$\pm$0.010} & 0.724{\tiny$\pm$0.011} & 0.720{\tiny$\pm$0.010} & 0.707{\tiny$\pm$0.011} \\
  & 5  & 0.630{\tiny$\pm$0.009} & 0.628{\tiny$\pm$0.011} & 0.630{\tiny$\pm$0.009} & 0.601{\tiny$\pm$0.010} & 0.734{\tiny$\pm$0.009} & 0.739{\tiny$\pm$0.011} & 0.734{\tiny$\pm$0.009} & 0.723{\tiny$\pm$0.010} \\
  & 6  & 0.643{\tiny$\pm$0.009} & 0.641{\tiny$\pm$0.010} & 0.643{\tiny$\pm$0.009} & 0.617{\tiny$\pm$0.010} & 0.745{\tiny$\pm$0.009} & 0.751{\tiny$\pm$0.010} & 0.745{\tiny$\pm$0.009} & 0.737{\tiny$\pm$0.010} \\
  & 7  & 0.650{\tiny$\pm$0.009} & 0.649{\tiny$\pm$0.010} & 0.650{\tiny$\pm$0.009} & 0.625{\tiny$\pm$0.010} & 0.756{\tiny$\pm$0.009} & 0.764{\tiny$\pm$0.010} & 0.756{\tiny$\pm$0.009} & 0.749{\tiny$\pm$0.009} \\
  & 8  & 0.659{\tiny$\pm$0.009} & 0.659{\tiny$\pm$0.010} & 0.659{\tiny$\pm$0.009} & 0.636{\tiny$\pm$0.010} & 0.766{\tiny$\pm$0.009} & 0.773{\tiny$\pm$0.010} & 0.766{\tiny$\pm$0.009} & 0.760{\tiny$\pm$0.009} \\
  & 9  & 0.666{\tiny$\pm$0.009} & 0.666{\tiny$\pm$0.010} & 0.666{\tiny$\pm$0.009} & 0.645{\tiny$\pm$0.010} & 0.772{\tiny$\pm$0.009} & 0.783{\tiny$\pm$0.009} & 0.772{\tiny$\pm$0.009} & 0.767{\tiny$\pm$0.010} \\
  & 10 & 0.673{\tiny$\pm$0.009} & 0.673{\tiny$\pm$0.010} & 0.673{\tiny$\pm$0.009} & 0.653{\tiny$\pm$0.010} & 0.776{\tiny$\pm$0.009} & 0.785{\tiny$\pm$0.010} & 0.776{\tiny$\pm$0.009} & 0.771{\tiny$\pm$0.010} \\
\bottomrule
\end{tabular}
}
\caption{Prediction Performance with Different Number of Actions and Shots. (Action Number 4--6)}
\end{table}

\end{document}